\documentclass[12pt, a4paper]{article}

\usepackage{epsfig}
\usepackage{graphicx}
\usepackage{multirow}
\usepackage{subfig}
\RequirePackage[colorlinks, citecolor=blue, urlcolor=blue]{hyperref}
\usepackage{natbib}
\usepackage{amsmath}
\usepackage{amsfonts}
\RequirePackage[OT1]{fontenc}

\newcommand{\bfh}{\hbox{\boldmath$h$}}

\newcommand{\bfp}{\hbox{\boldmath$p$}}
\newcommand{\bfr}{\hbox{\boldmath$r$}}
\newcommand{\bfs}{\hbox{\boldmath$s$}}
\newcommand{\bfx}{\hbox{\boldmath$x$}}

\newcommand{\bfalpha}{\hbox{\boldmath$\alpha$}}
\newcommand{\bfbeta}{\hbox{\boldmath$\beta$}}
\newcommand{\bftheta}{\hbox{\boldmath$\theta$}}
\newcommand{\bfpi}{\hbox{\boldmath$\pi$}}
\newcommand{\bfphi}{\hbox{\boldmath$\phi$}}
\newcommand{\bfGamma}{\hbox{\boldmath$\Gamma$}}
\newcommand{\bfPsi}{\hbox{\boldmath$\Psi$}}
\newcommand{\bftinyG}{\hbox{\tiny\textbf{G}}}
\newcommand{\bfPmat}{\hbox{\textbf{P}}}
\newcommand{\logit}{\hbox{logit}}

\begin{document}

\begin{center}
\Large {Estimating abundance from multiple sampling capture-recapture data via a multi-state multi-period stopover model} \\ \vspace{20pt}
\normalsize
Hannah Worthington$^1$, Rachel McCrea$^2$, Ruth King$^3$ and Richard Griffiths$^4$ \\ \vspace{10pt}
$^1$ School of Mathematics and Statistics, University of St Andrews, The Observatory, Buchanan Gardens, St Andrews, Fife, KY16 9LZ \\
$^2$ School of Mathematics, Statistics and Actuarial Science (SMSAS), University of Kent, Sibson Building, Parkwood Road, Canterbury, CT2 7FS \\
$^3$ School of Mathematics, University of Edinburgh, James Clerk Maxwell Building, The King's Buildings, Peter Guthrie Tait Road, Edinburgh, EH9 3FD \\ 
$^4$ Durrell Institute of Conservation and Ecology, School of Anthropology and Conservation, University of Kent, Marlowe Building, University of Kent, Canterbury, Kent, CT2 7NR 
\end{center}

\begin{abstract}
	The collection of capture-recapture data often involves collecting data on numerous capture occasions over a relatively short period of time.  For many study species this process is repeated, for example annually, resulting in capture information spanning multiple sampling periods.  The robust design class of models provide a convenient framework in which to analyse all of the available capture data in a single likelihood expression.  However, these models typically rely either upon the assumption of closure within a sampling period (the closed robust design) or condition on the number of individuals captured within a sampling period (the open robust design).  The models we develop in this paper require neither assumption by explicitly modelling the movement of individuals into the population both within and between the sampling periods, which in turn permits the estimation of abundance.  These models are further extended to allow parameters to depend not only on capture occasion but also the amount of time since joining the population and to the case of multi-state data where there is individual time-varying discrete covariate information.  We derive an efficient likelihood expression for the new multi-state multi-period stopover model using the hidden Markov model framework.  We demonstrate the new model through a simulation study before considering a dataset on great crested newts, \textit{Triturus cristatus}.
\end{abstract}


\section{Introduction} \label{sec:intro}
	In this paper we develop a model capable of analysing capture-recapture data from multiple sampling periods within a single likelihood expression.  In comparison to existing models we retain the ability to estimate population size through the likelihood and allow parameters to be dependent both on time and time spent in the population.  Standard capture-recapture studies consist of several capture occasions where attempts are made to capture individuals from the population of interest.  When an individual is captured for the first time it is marked, or unique physical marks recorded, to permit unique identification of each individual.  At subsequent capture occasions it is then possible, using these unique marks, to identify new individuals (which are subsequently marked) or recaptured individuals (those that have been previously captured).  In this paper, we assume that all sampled individuals are returned to the population after capture, i.e.~that there are no removals from the population.  By repeating this process at each capture occasion it is possible to identify on which occasions each unique individual was recorded.  This information is stored in the form of individual capture histories.  Typically these capture histories are of binary form, for example,
	\[
	0 \ 1 \ 1 \ 0 \ 0 \ 1 \ 0 \ 0
	\]
	where 0, and 1, indicate an individual was not captured, or captured, at each capture occasion respectively.  During some capture-recapture studies it may be possible to collect additional individual covariate information.  We consider the case where an individual time-varying discrete covariate is recorded corresponding to the state of the individual upon capture.  This additional information is recorded in the capture history where non zero entries now indicate the observed state.  This discrete state information may refer, for example, to behavioural states such as breeding or foraging, or alternatively it may refer to a discrete location such as the pond number on a field study site.
	
	The Cormack-Jolly-Seber (CJS) model \citep{Cormack64, Jolly65, Seber65} forms the basis of many commonly applied capture-recapture models.  Developed to estimate survival it conditions on the first capture of each individual and thus is unable to estimate the total population size.  To remove this assumption \citet{Schwarz96} proposed the idea of a super-population which includes both individuals that are captured at least once as well as those that are never captured (but are available for capture on at least one occasion).  The inclusion of the super-population in the Schwarz-Arnason (SA) model, denoted $N$, allows for births to be modelled within the likelihood expression.  The stopover model presented by \citet{Pledger09} is an extension of the SA model in which the capture and retention probabilities are dependent both on time and time since arrival.  In the stopover model the term `age' is used to refer to the time since joining the population (not necessarily physical age) and is generally unknown due to the unknown arrival time (an individual may have joined the population on an occasion before their first capture).  We note that when collecting data to which we wish to fit a stopover model it is advisable to sample both before the first arrivals and after the final departures (this results in capture histories with leading and trailing zeros).  Whilst the easiest approach is to analyse only the subset of data corresponding to when the site is occupied, the extended sampling before and after occupation can verify the implicit assumptions that those present on the first occasion a capture occurs have only recently arrived and those present on the final occasion where captures occur are imminently about to depart.
	
	Multi-state capture-recapture models extend these models to allow for individual time-varying discrete covariates.  For example, the Arnason-Schwarz (AS) model is a multi-state extension of the CJS model \citep{Arnason72, Arnason73, Brownie93, Schwarz93, King03, Lebreton09, King14b}.  \citet{Dupuis07} consider a similar multi-state extension of the SA model for estimating abundance in open populations fitted within a Bayesian (data augmentation) framework. This model allows for time- and state-dependence in the capture probabilities but  not the age-dependence of the stopover model.  Typically these models assume a first-order Markov model for the movement of individuals between the different discrete states.  \citet{King16} relax this assumption through a semi-Markov model where the dwell-time distribution (the time spent in the state) has some parametric form.
	
	The CJS, SA, stopover and AS models all consider a single group of capture occasions.  However, for many studies, capture occasions are spaced closely in time, for instance during the breeding season, and the sampling process is repeated many times, for example every year.  The robust design class of models consider the data at these two sampling levels; primary and secondary sampling periods.  In general the robust design models assume that the capture-recapture data of the secondary periods are collected over a relatively short period of time, whilst the duration between the primary sampling periods is much larger.  The closed robust design model \citep{Pollock82, Kendall95}, assumes that the population is open in the primary level, but closed in the secondary.  Such closed robust design models estimate abundance within each primary occasion.  However, the assumption of closure within the secondary sampling occasions can be unrealistic for many populations.  For example, some amphibians have breeding periods lasting a few weeks with each individual perhaps spending only one or two days at a breeding site.  The open robust design model \citep{Kendall01} retains the open primary occasions, but also permits the secondary occasions to be open to arrivals and departures.  However, these open robust design models cannot estimate abundance directly due to conditioning on the number of individuals captured in each primary occasion.
	
	We develop a general multi-state multi-period stopover model, and associated explicit likelihood expression, that extends the open robust design to a model capable of time- and age-dependence in the survival, capture and retention probabilities whilst also retaining the ability to estimate the total population size.  This new model retains the flexibility of movement into and out of the population, assuming an open population both between and within each sampling period, but without the need to condition on the number of individuals observed in each primary period.  We apply a similar argument to the stopover and SA models, assuming a total population across all the periods consisting of both those individuals that are observed and those that are not observed but available for capture.  This approach allows the size of the total population to be estimated, and subsequently the size of the population in each primary period.  The multi-state aspect of the model allows for additional information to be incorporated such as different mark types, location information or breeding status.  We focus in particular on allowing the capture probabilities to be state-dependent thus allowing for heterogeneity in the population.  This new multi-state multi-period stopover model can be considered a very general model for capture-recapture data from which all the existing models can all be obtained by placing appropriate restrictions on the model parameters.
	
	The motivation for developing this new multi-period stopover model is a long-term study on great crested newts, a protected species in Europe.  Although up to \pounds 43 million is spent on mitigating the impacts of development on this species in England alone \citep{Lewis17}, current population assessment protocols for this species are inadequate \citep{Griffiths15}.  There is consequently a need for more reliable statistical models that take account of the seasonal dynamics of this species.  The study population considered here is unique in that it is based  on replicated ponds that have been intensively monitored for nearly two decades.  Individuals in this population are captured weekly during the breeding season with the process repeated annually.  The additional state information for this population is the pond in which each individual newt is captured.  The field study site originally consisted of four ponds, a further four ponds were added in 2009 and first colonised in the 2010 breeding season. Given that pond creation is regarded as a fundamental component of amphibian conservation, of particular biological interest is how these new ponds have been colonised, whether capture probabilities differ between the old well-established ponds and the new ponds and the trap effectiveness at capturing the newts.  The old and new ponds may exhibit differences due to differing amounts of vegetation, with these differences perhaps disappearing as the new ponds become established.  For this population of newts there is particular interest in the total population size and the states themselves.
	
	For efficient evaluation of the likelihood we express the multi-state multi-period stopover model using a hidden Markov model (HMM) representation.  HMMs provide a flexible way of modelling series of observations collected through time that depend on underlying and often unobserved correlated states.  After the initial capture and marking of an individual, the capture history can be considered as a combination of two processes: the observation process which depends on the availability of an individual for capture; and an underlying availability process.  An HMM separates the underlying state process (i.e.~availability for capture) from the observation process (i.e.~capture process).  For further discussion see for example \citet{Gimenez07, Schofield08, Royle08, King09, King12, King14a, Langrock13, Zucchini16}.
	
	In Section~\ref{sec:model} we derive the multi-state multi-period stopover model.  In Section~\ref{sec:simstudy} we perform a simulation study before applying the new model to a data set on great crested newts in Section~\ref{sec:newts}.  We conclude with a discussion in Section~\ref{sec:discussion}.

\section{Model derivation} \label{sec:model}
	In this section we derive the multi-state multi-period stopover model.  When modelling age (time since arrival) in the stopover model we label each individual as being age 1 on the first occasion they attend the site to indicate that they have spent one capture occasion in the population.  In addition we use the general term `arrival' to indicate an individual becoming available for capture, this can in practice have different interpretations and could refer to births, recruitment to the breeding population or arrival at a specific colony for migratory species.  Likewise departures may refer to different ways of leaving a site, including deaths or permanent emigration from the study area.  In this derivation we incorporate the state-dependence in the capture probabilities and allow for movement between the states to be first-order Markov.  We also assume that the state of an individual is recorded without misclassification when an individual is observed, though this assumption can be relaxed \citep{King14b}.
	
	\subsection{Notation}
		In defining the notation of the multi-state multi-period stopover model we extend, where possible, the notation of \citet{Pledger09}.  Let $N$ denote the total population (to be estimated) consisting of all individuals who visit the study site for at least one capture occasion during the study period (all capture occasions and periods). Further, let $n$ denote the number of observed individuals (those captured on at least one capture occasion) and $n_m$ the number of individuals that are missed (those that are never captured).  Thus $N=n+n_m$.  Let the entire study period consist of $T$ primary periods, labelled $t = 1, \ldots, T$, with $K(t)$ secondary capture occasions in primary period $t$. We let the capture history for individual $i$ be denoted by $\bfx_i = \{x_i(t,k) : k = 1, \ldots, K(t); \ t = 1, \ldots, T\}$ and let the set of capture histories for all observed individuals be denoted by $\bfx = \{\bfx_i : i=1,\ldots,n\}$.  Note that from the histories we can easily extract in which primary periods each individual was captured at least once. 
	
		We now define the set of model parameters (in addition to $N$ above).  We define the recruitment probabilities to be the set $\bfr = \{r(t) : t = 1, \ldots, T\}$ where $r(t)$ is the probability of being recruited into the population and first becoming available for capture in primary period $t$. Since an individual belonging to the total population must visit the site during at least one primary period, $\sum_{t=1}^{T}{r(t)} = 1$. For the HMM formulation of the model we define $r^*(t) = r(t) / \sum_{ j=t}^{T}{r(j)}$ for $t=2,\ldots,T$ which denotes the conditional recruitment probability (probability of being recruited in primary period $t$ given the individual has not been recruited in any primary periods $1,\ldots,t-1$).  We define the set of arrival probabilities to be $\bfbeta = \{\beta(t,k): k=1,\ldots,K(t); \ t=1,\ldots,T\}$ where $\beta(t,k)$ is the probability of arriving at the study site and being available for capture from occasion $k$ within primary period $t$, given the individual is in the population and available for capture in primary period $t$.  By definition, within each primary period $t=1,\ldots,T$, $\sum_{k=1}^{K(t)}{\beta(t,k)} = 1$.  Similarly to the recruitment probabilities, the HMM formulation requires conditional arrival probabilities which we define as $\beta^*(t,k) = \beta(t,k) / \sum_{j=k}^{K(t)}{ \beta(t,j)}$ for $k=2,\ldots,K(t)$ and $t=1,\ldots,T$ (probability of arriving on occasion $k$ in primary period $t$ given the individual has not arrived on occasions $1,\ldots,k-1$ in primary period $t$).  We let $\bfs = \{s_A(t): A=1,\ldots,t; \ t=1,\ldots,T-1\}$ denote the set of survival probabilities, where $s_A(t)$ is the probability an individual is available for capture in primary period $t+1$ given they are `Age' $A$ (have been present in the population for $A$ primary periods) and available for capture in primary period $t$.  We let $\bfphi = \{\phi_a(t,k): a=1,\ldots,k; \ k=1,\ldots,K(t)-1; \ t=1,\ldots,T\}$ denote the set of retention probabilities where $\phi_a(t,k)$ is the probability that an individual is available for capture on occasion $k+1$ in primary period $t$ given the individual is age $a$ (has been present in the population for $a$ secondary period capture occasions within primary period $t$) and available for capture on occasion $k$ in primary period $t$.  
	
		In order to model the movement of individuals between the different observable discrete states we first need to consider the discrete state that an individual enters when they first arrive at the site within each primary period.  We denote these initial discrete state probabilities by $\bfalpha = \{\alpha_g(t): t=1,\ldots,T; \ g=1,\ldots,G\}$ where $\alpha_g(t)$ is the probability of being in state $g=1,\ldots,G$ (where $G$ is the total number of observable states) on the first occasion an individual is available for capture in primary period $t$.  In this derivation we assume these initial discrete state probabilities are constant over time and so regardless of when an individual arrives at the site the probability they enter each of the observable states remains the same.   
	
		The set of transition probability matrices between the discrete states is denoted by $\bfPsi = \{\bfPsi(t):t=1,\ldots,T\}$.  The transition probabilities in primary period $t$ are given by,
		\begin{align*}
			\bfPsi(t) & = \left(\begin{array}{cccc}
				\psi_{11}(t) & \psi_{12}(t) & \ldots & \psi_{1G}(t)\\
				\psi_{21}(t) & \psi_{22}(t) & \ldots & \psi_{2G}(t)\\
				\vdots & \vdots & \ddots & \vdots\\
				\psi_{G1}(t) & \psi_{G2}(t) & \ldots & \psi_{GG}(t)\\
			\end{array} \right)
		\end{align*}
		such that $\psi_{ij}(t)$ denotes the probability of moving from state $i$ to state $j$ between consecutive secondary occasions in primary period $t$, conditional on the individual remaining available for capture in primary period $t$.  For simplicity we have defined the transition probabilities to be constant across all occasions within a primary period.  In general, this need not be the case, however, there are likely to be issues with parameter redundancy and identifiability in the fully time-dependent case.  Finally, we define the capture probabilities to be $\bfp = \{p_{ga}(t,k): a=1,\ldots,k; \ k=1,\ldots,K(t); \ t=1,\ldots,T; \ g=1,\ldots,G\}$ where $p_{ga}(t,k)$ is the probability an individual is captured given they are in state $g$ and age $a$ on occasion $k$ in primary period $t$.  The full set of model parameters for the multi-state multi-period stopover model is given by $\bftheta = \{N, \bfr, \bfs, \bfalpha, \bfPsi, \bfbeta, \bfphi, \bfp\}$.
	
	\subsection{HMM formulation}
		Following the convention of the robust design models we consider nested (or hierarchical) Markov chains, the first operating on the primary level and the second nested chain operating on the secondary capture occasions.  Let $\bfh = \{h(t): t=1,\ldots,T\}$ be the hidden states in the primary level where,
		\begin{align*}
			h(t) & = \left\{ \begin{array}{ll}
				1 & \hbox{not yet recruited into the attending population;}\\
				2 & \hbox{Age 1 in the attending population;}\\
				\vdots & \vdots\\
				A'+1 & \hbox{Age $A'$ in the attending population;}\\
				A'+2 & \hbox{departed from the attending population;}\\
			\end{array} \right.
		\end{align*}
		where $A'$ is the maximum observable age of individuals in the population on the primary level $(A' \leq T)$.  Similarly, let $\bfh(t) = \{h(t,k): k=1,\ldots,K(t); \ t=1,\ldots,T\}$ be the hidden states in the secondary level where,
		\begin{align*}
			h(t,k) & = \left\{\begin{array}{ll}
				1 & \hbox{not yet available for capture;}\\
				2 & \hbox{available for capture in primary $t$, age 1 and in state 1;}\\
				3 & \hbox{available for capture in primary $t$, age 1 and in state 2;}\\
				\vdots & \vdots\\
				G+1 & \hbox{available for capture in primary $t$, age 1 and in state $G$;}\\
				G+2 & \hbox{available for capture in primary $t$, age 2 and in state 1;}\\
				\vdots & \vdots\\
				2G+1 & \hbox{available for capture in primary $t$, age 2 and in state $G$;}\\
				\vdots & \vdots\\
				a'(t) G+1 & \hbox{available for capture in primary $t$, age $a'(t)$ and in state $G$;}\\
				a'(t) G+2 & \hbox{departed from the site in primary $t$;}\\
			\end{array} \right.
		\end{align*}
		where $a'(t)$ is the maximum observable age of individuals in the secondary level $(a'(t) \leq K(t))$.  We note that it is possible that $a'(t)$ and $G$ could be different in each primary period.  This would change the size of the matrices used within the secondary level of the model but no other changes are necessary. We also note that the age need not increment by one each time but could more generally refer to age classes, for example, immature, adult and senior. Let the initial hidden state distribution of the primary level HMM, 
		$$\bfpi(1) = \left(\begin{array}{cccc}
			\mathbb{P}(h(1)=1) & \mathbb{P}(h(1)=2) & \ldots & \mathbb{P}(h(1)=A'+2)
		\end{array} \right)$$
		be the probabilities of entering each primary hidden state for primary period 1.  Similarly, for the secondary level HMM, 
		$$\bfpi(t,1) = \left(\begin{array}{cccc}
			(\mathbb{P}(h(t,1)=1) & \mathbb{P}(h(t,1)=2) & \ldots & \mathbb{P}(h(t,1)=a'(t)G+2))
		\end{array} \right)$$
		for $t=1,\ldots,T$ describes the probabilities of entering each secondary hidden state on occasion 1 in each primary period $t=1,\ldots,T$.  Then, by definition of the model parameters above,
		\begin{align*}
			\bfpi(1) & = \left(\begin{array}{ccccc}
				1-r(1) & r(1) & 0 & \ldots & 0
			\end{array} \right)\\
			\bfpi(t,1) & = \left( \begin{array}{ccccc}
				1-\beta(t,1) & \beta(t,1) \bfalpha(t) & 0 & \ldots & 0
			\end{array} \right)
		\end{align*}
		where $\bfalpha(t)=\left(\begin{array}{cccc} \alpha_1(t) & \alpha_2(t) & \ldots & \alpha_G(t) \end{array} \right)$ is the set of initial discrete state probabilities for primary period $t$.  Next we consider the transition matrices which describe the movement between the states of the Markov chains.  In the primary level this concerns the survival between the primary periods whilst in the secondary level it is the retention within the given primary period.  Let $\bfGamma(t)$ be an $(A'+2)\times (A'+2)$ matrix where $$\bfGamma(t)[a,b] = \mathbb{P}(h(t+1)=b | h(t)=a)$$
		for $t=1,\ldots,T-1$, $a=1,\ldots,A'+2$ and $b=1,\ldots,A'+2$.  Similarly, let $\bfGamma(t,k)$ be an $(a'(t)G+2)\times (a'(t)G+2)$ matrix where 
		$$\bfGamma(t,k)[a,b] = \mathbb{P}(h(t,k+1)=b|h(t,k)=a)$$
		for $k=1,\ldots,K(t)-1$, $t=1,\ldots,T$, $a=1,\ldots,a'(t)G+2$ and $b=1,\ldots,a'(t)G+2$.  By definition,
		\scriptsize
		\begin{align*}
			\bfGamma(t) & = \left( \begin{array}{ccccccc}
				1-r^*(t+1) & r^*(t+1) & 0 & 0 & \ldots & 0 & 0\\
				0 & 0 & s_1(t) & 0 & \ldots & 0 & 1-s_1(t)\\
				0 & 0 & 0 & s_2(t) & \ldots & 0 & 1-s_2(t)\\
				\vdots & \vdots & \vdots & \vdots & \ddots & \vdots & \vdots\\
				0 & 0 & 0 & 0 & \ldots & s_{A'-1}(t) & 1-s_{A'-1}(t)\\
				0 & 0 & 0 & 0 & \ldots & 0 & 1\\
				0 & 0 & 0 & 0 & \ldots & 0 & 1\\
			\end{array} \right)
		\end{align*}
		\begin{align*}
			\bfGamma(t,k) & = \left(\begin{array}{cccccc}
				1-\beta^*(t,k+1) & \beta^*(t,k+1) \bfalpha(t) & 0 & \ldots & 0 & 0\\
				0 & 0 & \phi_1(t,k) \bfPsi(t) & \ldots & 0 & (1-\phi_1(t,k))_{\bftinyG}\\
				\vdots & \vdots & \vdots & \ddots & \vdots & \vdots\\
				0 & 0 & 0 & \ldots & \phi_{a'(t)-1}(t,k) \bfPsi(t) & (1-\phi_{a'(t)-1}(t,k))_{\bftinyG}\\
				0 & 0 & 0 & \ldots & 0 & \bf1_G\\
				0 & 0 & 0 & \ldots & 0 & 1\\
			\end{array} \right)
		\end{align*}
		\normalsize
		where $(1-\phi_a(t,k))_{\bftinyG}$ is a column vector of length $G$ with each entry equal to $(1-\phi_a(t,k))$ and $\bf1_G$ is a column vector of ones of length $G$. 
	
		Finally we consider the observation process which connects the observed data to the hidden states.  The primary level relates to the probability of observing the capture histories within each primary period and the secondary level relates to the probability of capture on each occasion.  We first consider the secondary level and work with unique capture histories $j=1,\ldots,J$ rather than considering each individual in turn ($J \leq n$).  Let $\bfPmat(t,k,x_j(t,k))$ be an $(a'(t)G+2)\times (a'(t)G+2)$ diagonal matrix for $k=1,\ldots,K(t)$ and $t=1,\ldots,T$ where $\bfPmat(t,k,a)[b,b] = \mathbb{P}(x_j(t,k)=a | h(t,k)=b)$ for $a=0,1,\ldots,G$ and $b=1,\ldots,a'(t)G+2$ and all off-diagonal entries are zero.  Then,
		\small
		\begin{align*}
			\bfPmat(t,k,x_j(t,k)) & = \left\{\begin{array}{ll}
				\hbox{diag}(1,1-p_{11}(t,k),1-p_{21}(t,k),\ldots,1-p_{G1}(t,k), & \\
				\hspace{1.5cm} \ldots,1-p_{1a'(t)}(t,k),\ldots,1-p_{Ga'(t)}(t,k),1) & x_j(t,k) = 0\\
				\hbox{diag}(0,p_{11}(t,k),0,\ldots,0,p_{12}(t,k),0, & \\
				\hspace{1.5cm} \ldots,0,p_{1a'(t)}(t,k),0,\ldots,0) & x_j(t,k) = 1\\
				\hbox{diag}(0,0,p_{21}(t,k),0,\ldots,0,p_{22}(t,k),0, & \\
				\hspace{1.5cm} \ldots,0,p_{2a'(t)}(t,k),0,\ldots,0) & x_j(t,k) = 2\\
				\vdots & \vdots\\
				\hbox{diag}(0,\ldots,0,p_{G1}(t,k),0,\ldots,0,p_{G2}(t,k),0, & \\
				\hspace{1.5cm} \ldots,0,p_{Ga'(t)}(t,k),0) & x_j(t,k) = G.\\
			\end{array} \right.
		\end{align*}
	
		\normalsize
		Let $L_0(t)$ and $L_j(t)$ denote the likelihood contribution for a single-period stopover model (i.e.~considering the secondary occasions within one primary occasion only) for an all zero capture history (i.e.~an individual that is not captured) or a non-zero capture history in primary period $t$ respectively. Then for each primary period $t=1,\ldots,T$,
		\begin{align*}
			L_0(t) & = \bfpi(t,1) \bfPmat(t,1,0) \left( \prod_{k=2}^{K(t)}{ \bfGamma(t,k-1) \bfPmat(t,k,0)} \right) \bf1_{a'(t)G+2}\\
			L_j(t) & = \bfpi(t,1) \bfPmat(t,1,x_j(t,1)) \left( \prod_{k=2}^{K(t)}{ \bfGamma(t,k-1) \bfPmat(t,k,x_j(t,k))} \right) \bf1_{a'(t)G+2}
		\end{align*}
		where $\bf1_{a'(t)G+2}$ is a column of ones of length $a'(t)G+2$ (the number of states in the secondary level of the HMM).  We can now consider the observation process in the primary level.  Let $z_j(t)$ indicate whether capture history $j=1,\ldots,J$  contains a capture in primary period $t$.  Then $z_j(t) = 0$ if $x_j(t,k) = 0$ for all $k=1,\ldots,K(t)$ and conversely $z_j(t)=1$ if $x_j(t,k) \neq 0$ for at least one occasion $k=1,\ldots,K(t)$.  Let $\bfPmat(t,z_j(t))$ be an $(A'+2)\times (A'+2)$ diagonal matrix for $t=1,\ldots,T$ where $\bfPmat(t,a)[b,b] = \mathbb{P}(z_j(t)=a | h(t)=b)$ for $a=0,1$ and $b=1\ldots,A'+2$ and all off-diagonal entries are zero.  Then,
		\begin{align*}
			\bfPmat(t,z_j(t)) & = \left\{ \begin{array}{ll}
				\hbox{diag}(1,L_0(t),\dots,L_0(t),1) & z_j(t)=0\\
				\hbox{diag}(0,L_j(t),\dots,L_j(t),0) & z_j(t)=1.
			\end{array} \right.
		\end{align*}
	
		Let $L_0$ denote the probability an individual is never captured and $L_j$ the probability of observing the unique (non-zero) capture history $j$, then the primary level expressions for the HMM are,
		\begin{align*}
			L_0 & = \bfpi(1) \bfPmat(1,0) \left( \prod_{t=2}^{T}{ \bfGamma(t-1) \bfPmat(t,0)} \right) \bf1_{A'+2}\\
			L_j & = \bfpi(1) \bfPmat(1,z_j(1)) \left( \prod_{t=2}^{T}{ \bfGamma(t-1) \bfPmat(t,z_j(t))} \right) \bf1_{A'+2}
		\end{align*}
		where $\bf1_{A'+2}$ is a column of ones of length $A'+2$ (the number of states in the primary level of the HMM).
	
		The expression for the full likelihood is of multinomial form where individuals with the same capture history are grouped together.  We let $n_j$ denote the frequency of each unique capture history $j=1,\ldots,J$ where $J$ is the total number of unique non-zero capture histories.  The likelihood expression is given by: 
		\begin{align*}
			L(\bftheta | \bfx) & = \frac{N!}{ (N-n)! \prod_{j=1}^{J}{n_j!}} L_0^{N-n} \prod_{j=1}^{J}{L_j^{n_j}}.
		\end{align*}
	
		Thus we have an explicit likelihood expression.

\section{Simulation study} \label{sec:simstudy}
	To demonstrate the ability to estimate the parameters of the multi-state multi-period stopover model we perform a simulation study.  To explore the advantages of the new approach we compare the results of fitting a multi-state multi-period stopover model against the results of fitting separate multi-state stopover models independently to each primary period of data.
	
	We consider two different total population sizes, $N=100$ and $N=1000$, to determine the effect of population size on the ability to estimate the model parameters.  We expect that for small total population sizes (where the number of individuals captured in any one primary period will be relatively small) the multi-period model will perform better than the single-period models by taking strength from sharing parameters across the different primary periods.  As the population size increases we expect the variation of parameter estimates to decrease and the performance of the single-period models to improve.
	
	We generate three primary periods of data $(T=3)$ with each primary period having five capture occasions ($K(t)=5$ for all $t=1,\ldots,T$).  We let the number of individuals joining the population at each primary period follow a multinomial distribution with probabilities $r(1)=0.4$, $r(2)=0.2$ and $r(3)=0.4$. The probability of survival between each primary period is assumed to be constant with value $s=0.7$.  To model the arrivals within each primary period we define a logistic regression with a primary-dependent gradient, such that
	\begin{align*}
		\logit(\beta(t,k)) & = (\eta(t) k + \delta) \times \frac{1}{\logit(\beta(t,K(t))}
	\end{align*}
	where $\eta(1)=-1$, $\eta(2)=0$, $\eta(3)=-2$ and $\delta=1$.  The division by $\logit(\beta(t,K(t))$ ensures the arrival probabilities sum to one. For the retention probabilities we also use a logistic regression which is the same for each primary period and contains time effects and a linear age term so that,
	\begin{align*}
		\logit(\phi_a(t,k)) & = \tau(k) + \gamma(a-1)
	\end{align*}
	where $\tau(1)=2.5$, $\tau(2)=1.8$, $\tau(3)=2.1$, $\tau(4)=1.4$ and $\gamma=-1$.  
	
	For the state-dependent parameters we assume them to be constant across all primary occasions.  The initial discrete state probabilities are $\alpha_1=0.35$ and so $\alpha_2=0.65$.  The capture probabilities we assume to be dependent on state only with $p_1=0.6$ and $p_2=0.8$.  Finally we let the transition probability matrix between the observable discrete states be
	\begin{align*}
		\bfPsi & = \left(\begin{array}{cc}
			0.4 & 0.6\\
			0.3 & 0.7\\
		\end{array} \right).
	\end{align*}
	
\begin{figure}[!htp]
		\begin{center}
			\includegraphics[width=0.45\textwidth]{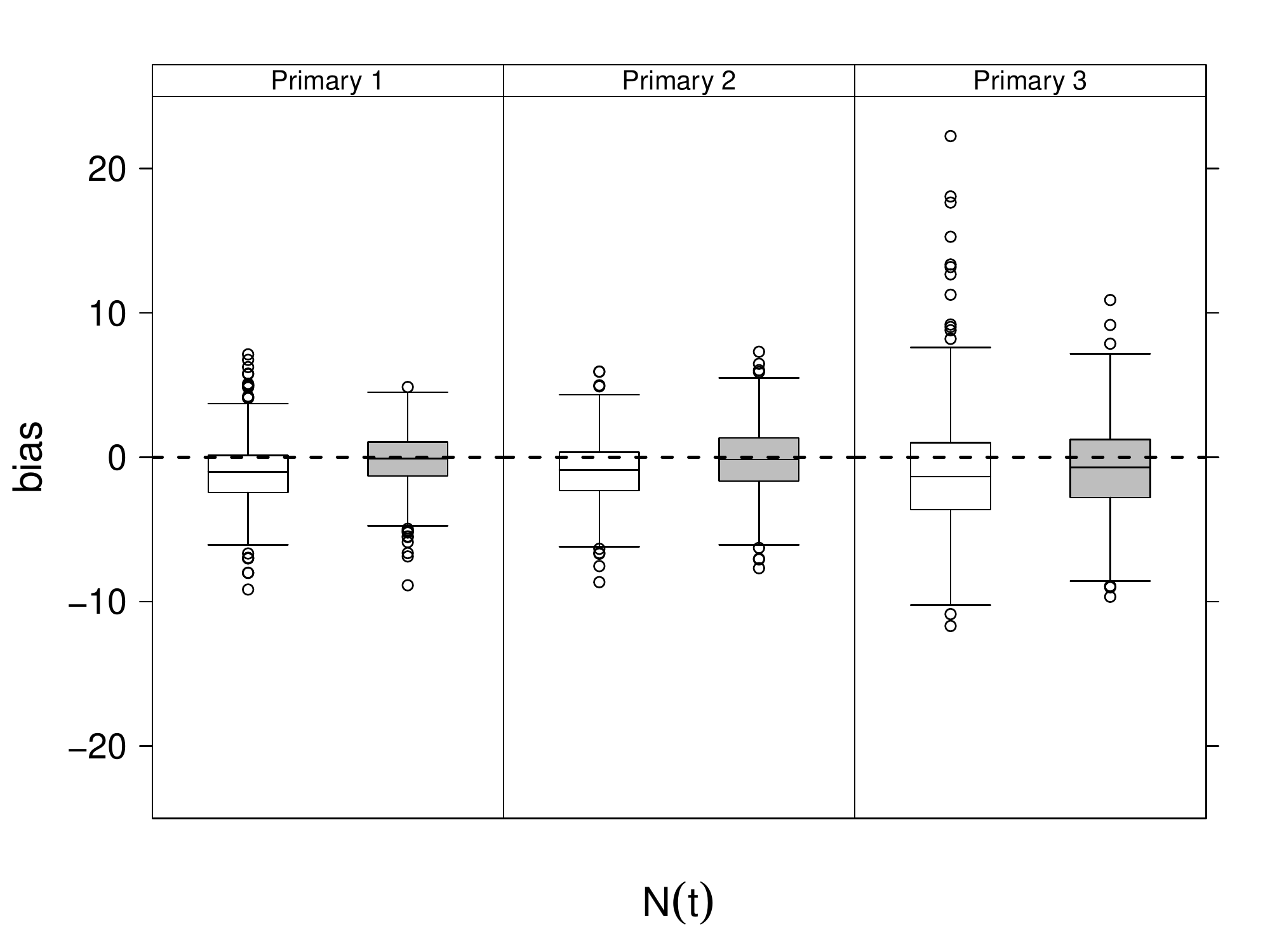}
			\includegraphics[width=0.45\textwidth]{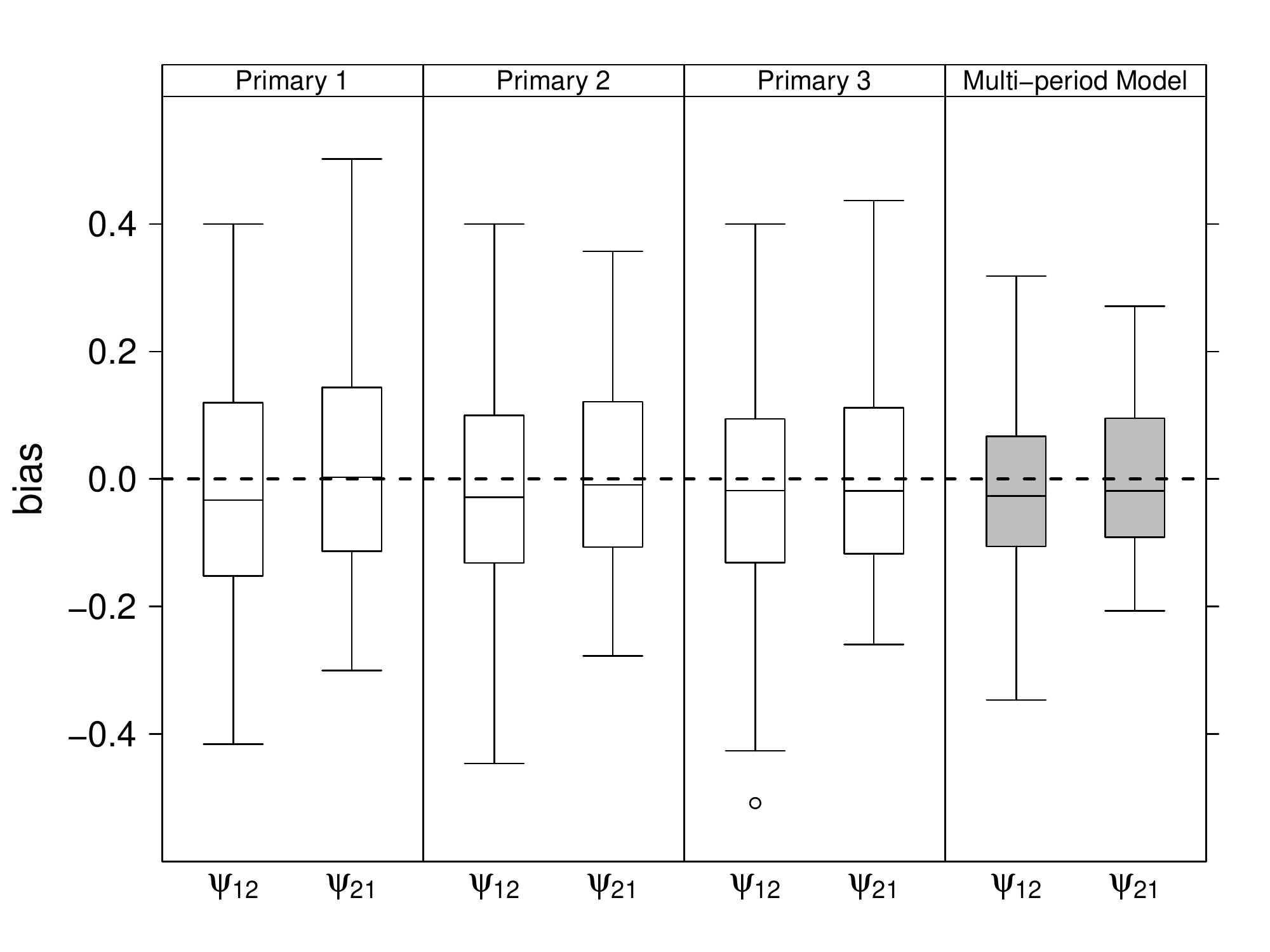}
			\includegraphics[width=0.45\textwidth]{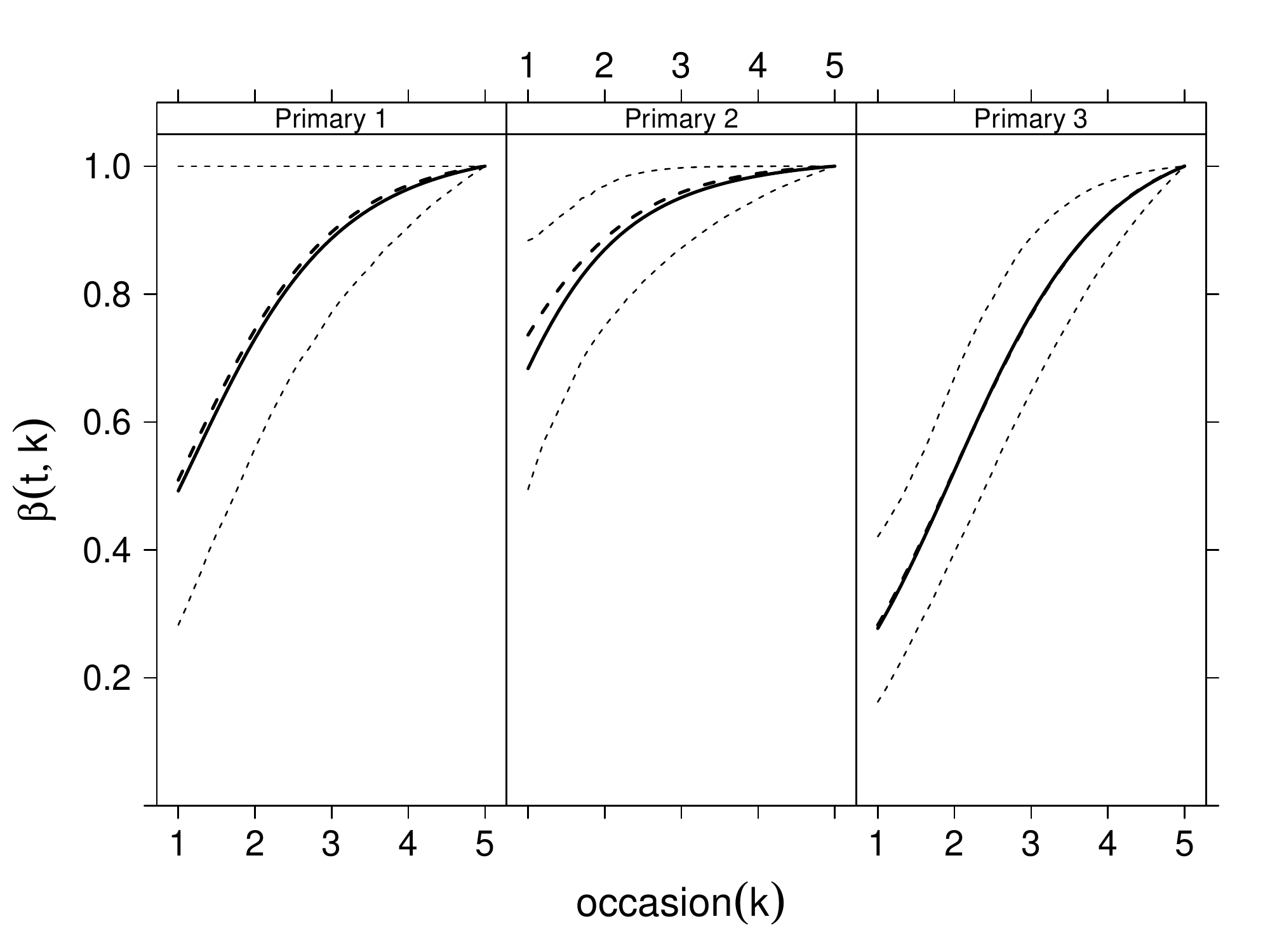}
			\includegraphics[width=0.45\textwidth]{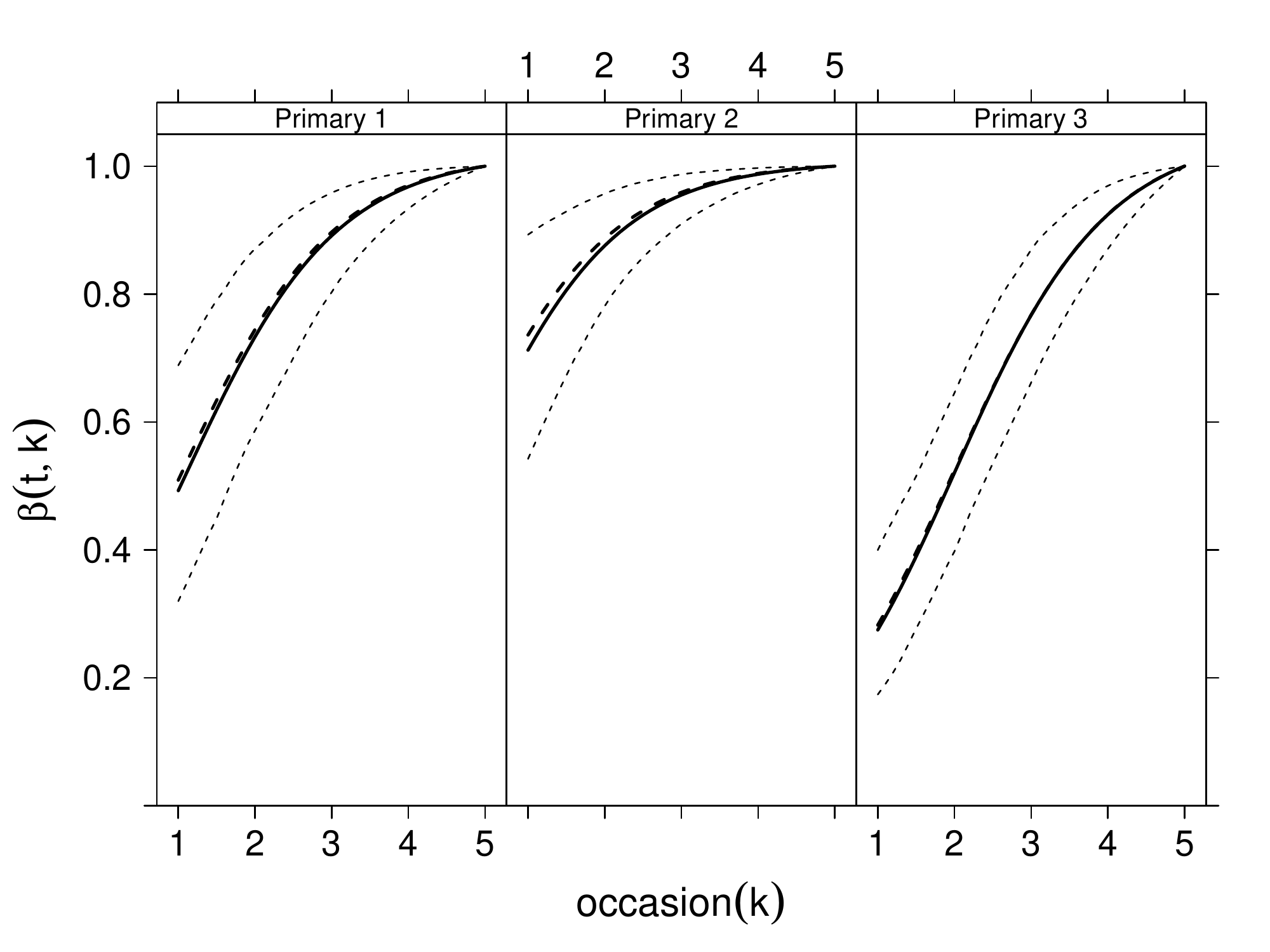}
		\end{center}
		\caption{Results from the simulation study where $N=100$ for: (top, left) bias of the population size estimates in each primary period for the single-period model (white) and multi-period model (grey); (top, right) bias of the transition probabilities in each primary period for the single-period and multi-period model; (bottom, left) logistic regression of the arrival probability (average across all data sets) for each primary period for the single-period model with 95\% percentile intervals and; (bottom, right) logistic regression of the arrival probability (average across all data sets) for each primary period for the multi-period model with 95\% percentile intervals.  True parameter values are shown by a dashed line.}
		\label{fig:N100ms}
	\end{figure}

	For each population size we generate 1000 data sets.  For each data set we fit the multi-state single-period stopover model to each year of data and then the multi-state multi-period stopover model to the full data set using the \texttt{nlm} function in \texttt{R} to maximise the likelihoods.  The results for population size $N(t)$ (estimated using the MLEs for $N$, $\bfr$ and $s$ for the multi-period model), transition probabilities $\bfPsi$ and the arrival probabilities $\bfbeta$ are displayed in Figures~\ref{fig:N100ms} and \ref{fig:N1000ms} for $N=100$ and $N=1000$ respectively (output for the remaining parameters is available in Supplementary Appendix A). 
	
	\begin{figure}[!htp]
		\begin{center}
			\includegraphics[width=0.45\textwidth]{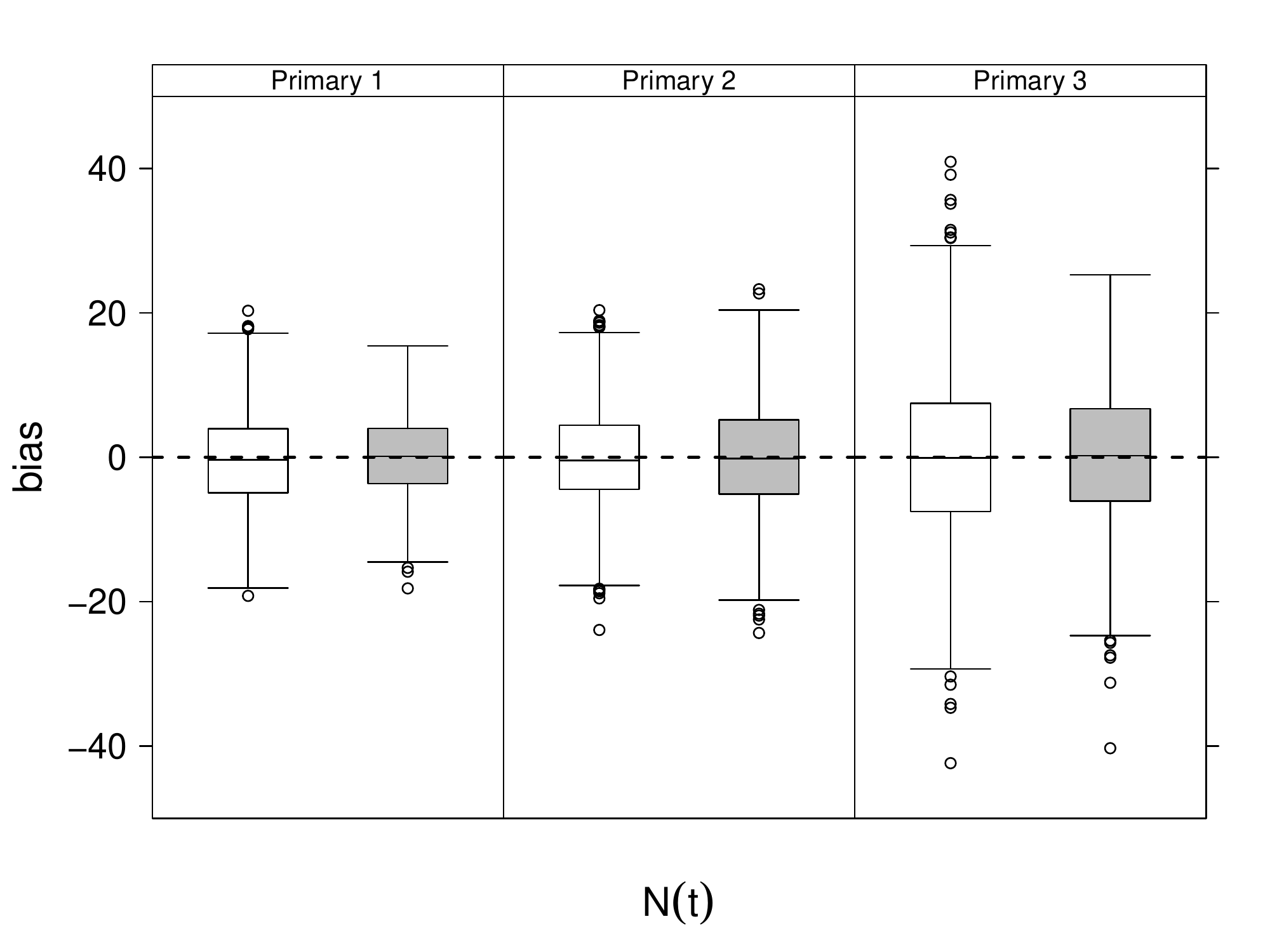}
			\includegraphics[width=0.45\textwidth]{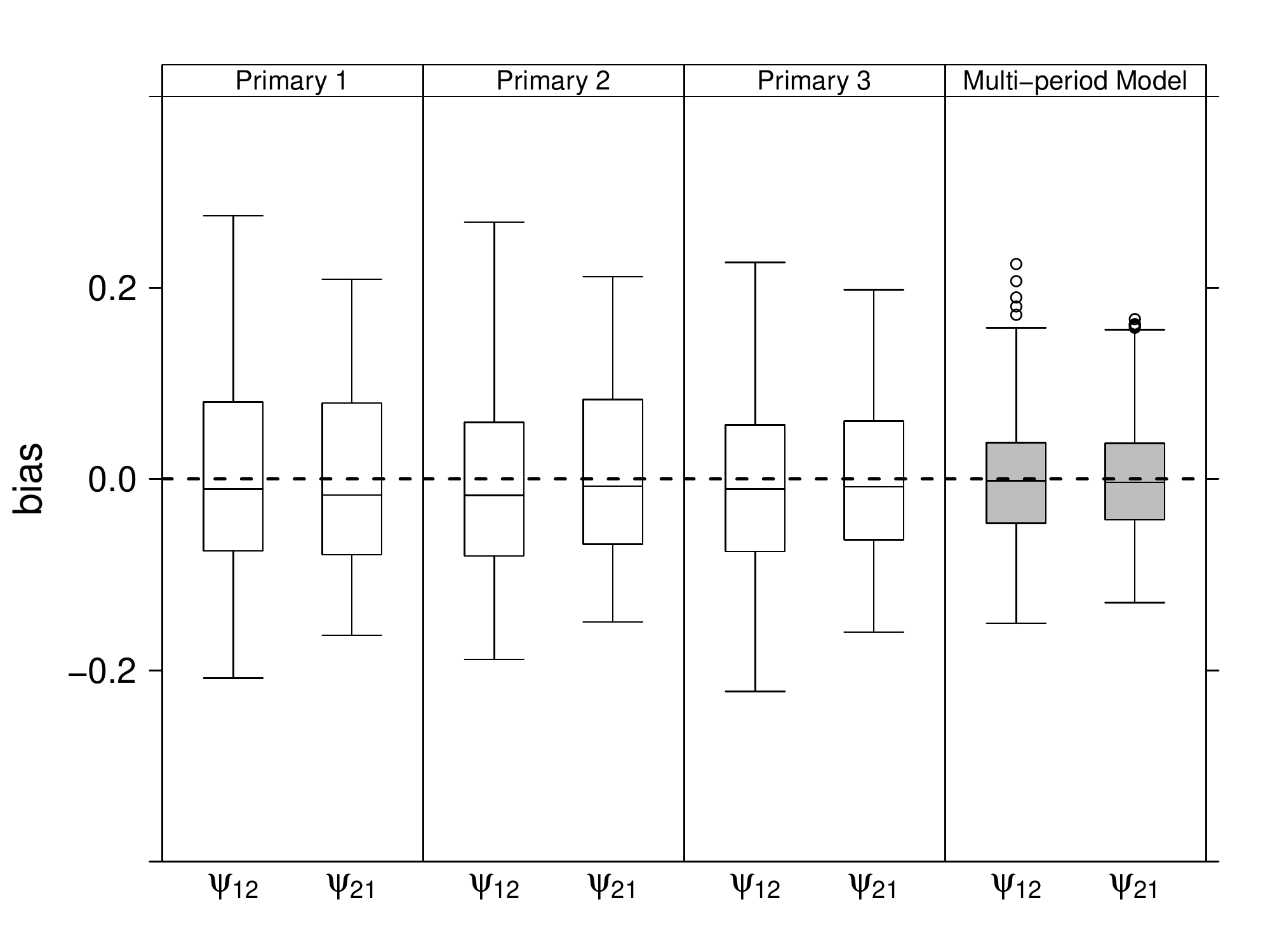}
			\includegraphics[width=0.45\textwidth]{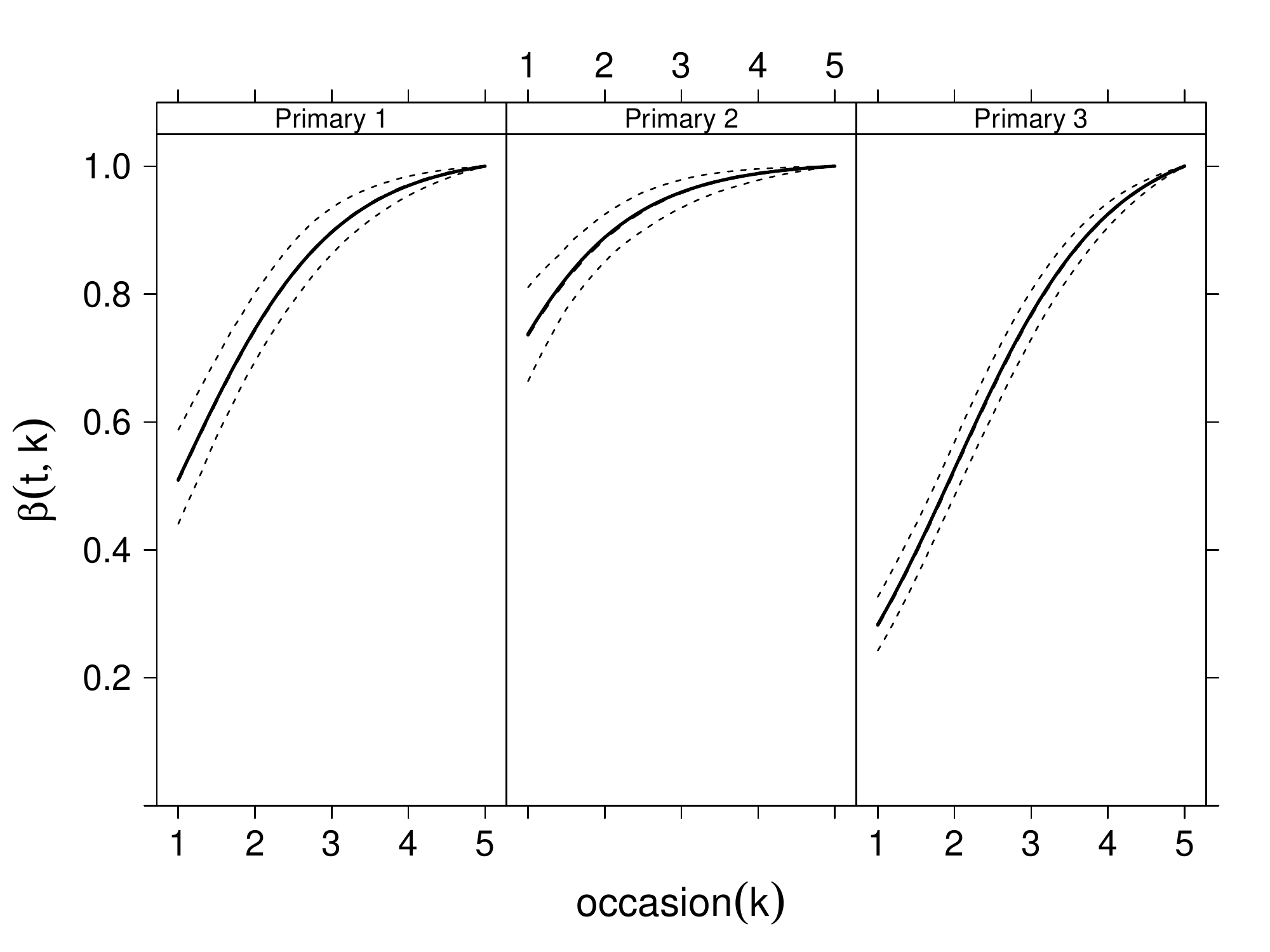}
			\includegraphics[width=0.45\textwidth]{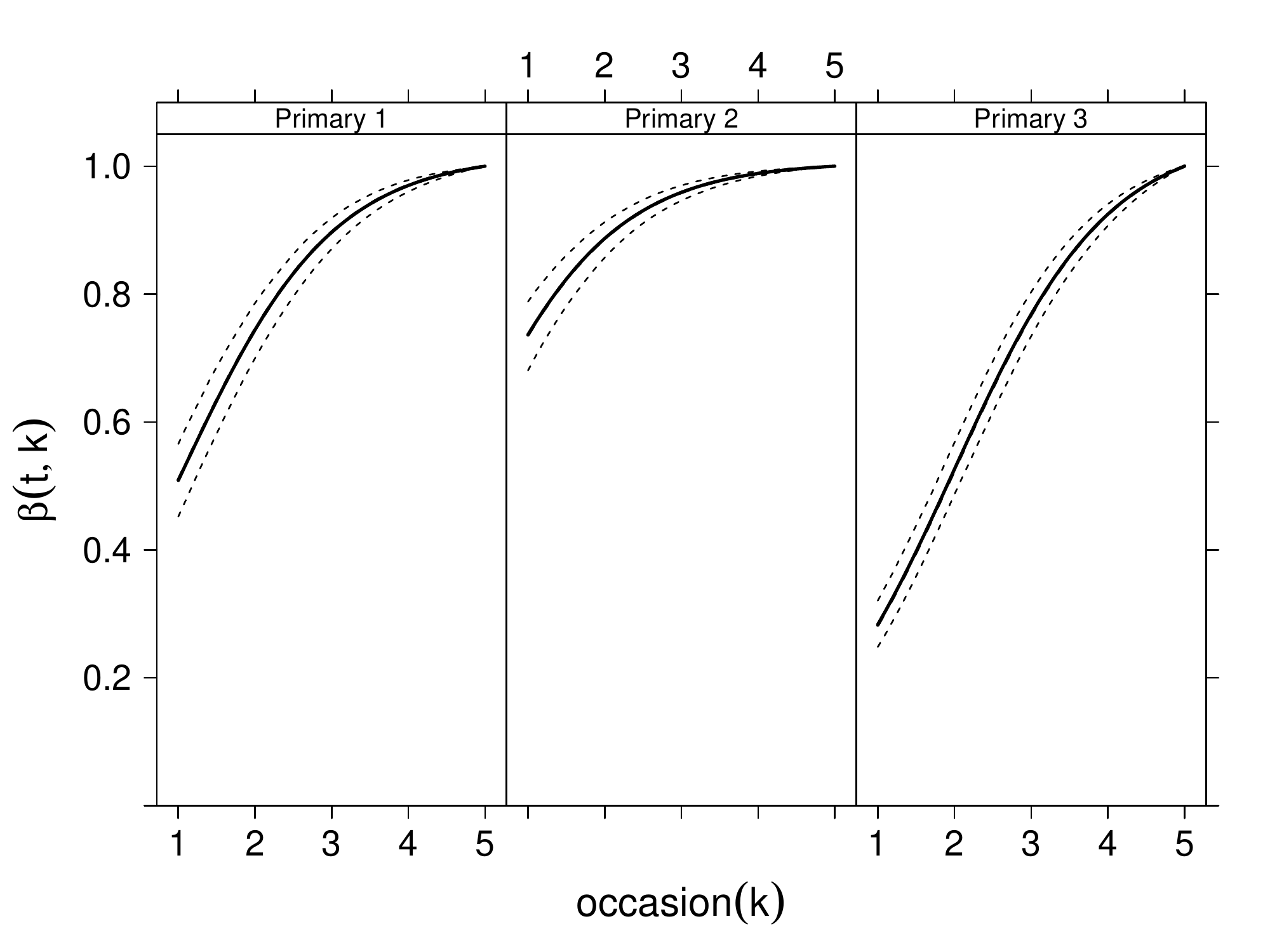}
		\end{center}
		\caption{Results from the simulation study where $N=1000$ for: (top, left) bias of the population size estimates in each primary period for the single-period model (white) and multi-period model (grey); (top, right) bias of the transition probabilities in each primary period for the single-period and multi-period model; (bottom, left) logistic regression of the arrival probability (average across all data sets) for each primary period for the single-period model with 95\% percentile intervals and; (bottom, right) logistic regression of the arrival probability (average across all data sets) for each  primary period for the multi-period model with 95\% percentile intervals.  True parameter values are shown by a dashed line.}
		\label{fig:N1000ms}
	\end{figure}
	
	From the simulation study we can clearly see the improved performance through using the multi-period approach.  All of the parameters are estimated well and appear to be unbiased (or close to unbiased).  We particularly note the improvement in the bias of the estimates for the population size in each primary period and the decrease in variability of the transition and arrival probabilities.  Of particular interest is the ability of the models to correctly estimate the state-dependent parameters.  For these, when $N=100$, the multi-period model does provide lower variability in the MLEs than the single-period approach though uncertainty in these parameters is quite large compared to the other parameters in the model.  Similar improvements to the precision of transition parameters were found by \citet{McCrea10} where a multi-state integrated population modelling approach is used to jointly analyse capture-recapture and census data.  We also note the strong improvement in the estimation of the retention probabilities ($\bfphi$) using the multi-period approach.
	
	When the total population size is increased to $N=1000$ the variation in estimates for all parameters in both models is reduced.  The multi-period model still performs better than the single-period model with the most obvious improvement now in the capture, initial discrete state and transition probabilities.  The variation in parameter $p_1$ in the multi-period model remains greater than the other parameters, this is most likely due to the low capture probability and the probability of remaining in the state being 0.4.  This is resulting in only a small number of captures in this state and so larger uncertainty than the equivalent parameter in the other state.  However, we do note that the estimates are unbiased and so in general the parameters appear to be identified well.

\section{Application} \label{sec:newts}
	Data on a population of great crested newts are collected from a field study site on the University of Kent campus.  The data have been collected since 2002 on a weekly basis throughout the breeding season following a standard and repeatable sampling routine.  Whilst all captured newts are recorded we analyse only the adult newt data since the natural markings used to uniquely identify individuals may still be developing in juvenile newts.  We consider the data collected between 2002-2013 inclusive, a total of 12 years of data.  In total there are 253 capture occasions across the 12 years.  The number of capture occasions each year varies; traps are set from the final week in February, which is typically before any newts arrive, and continue to be set until no further newts are captured or the water level in the ponds falls making trapping problematic.  We format the data such that the first capture occasion occurs within the same week every year (this may require truncating leading zeros from the capture histories within some years). Originally consisting of four ponds the site was extended in 2009 to a total of eight ponds which were then first colonised during the 2010 breeding season.  We define the observable capture states to be the type of pond (old or new) the individual is captured in; the `old' ponds were available in all years 2002-2013 whilst the second state, `new' ponds, were available in years 2010-2013.  The ponds are all located close together at the field study site (1-12m apart) and so movement between all eight ponds is possible and it is the environmental differences (for example the amount of vegetation) between the old and the new ponds that is likely to affect the choice of pond.  In total $n=106$ unique individuals were captured during the 12 years of sampling.
	
	To consider the choice of model we first model the capture-recapture data, without considering the additional state information, using the HMM formulation of the multi-period stopover model.  We perform a systematic search through a series of models of varying complexity in terms of the parameter dependencies.  We start with the most basic model where all the parameters are considered to be constant and shared across all years.  Improvement in the model fit is determined through the AIC statistic using a `step-up' approach in order to avoid choosing an overly complex model \citep{McCrea11}.  Due to the large number of capture occasions, we use a logistic regression on both the arrival and retention probabilities (rather than estimating probabilities for each occasion separately, this approach would require a very large number of parameters and the sample size here is comparatively small).  The model chosen by AIC (where the state information is ignored) includes year-dependent recruitment probabilities, a constant survival probability between each breeding season, and capture probabilities that are both year- and occasion-dependent i.e.~a different capture probability on every capture occasion.  For the logistic regressions on arrival and retention, the intercepts are constant and shared across all years whilst the gradients are year-dependent, with the gradient estimated separately for each year.
	
\begin{figure}[!htp]
		\begin{center}
			\includegraphics[width=0.45\textwidth]{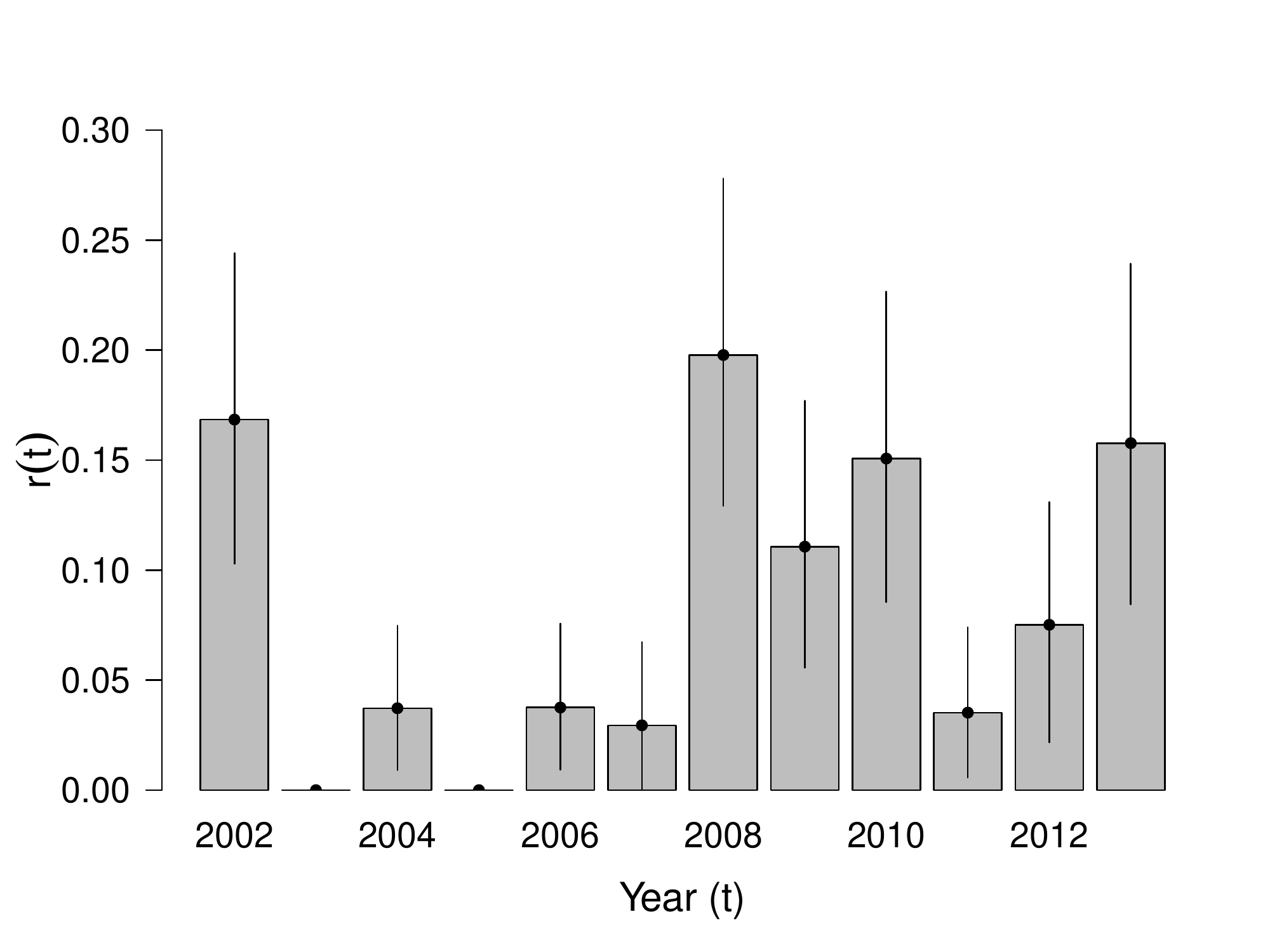}
			\includegraphics[width=0.45\textwidth]{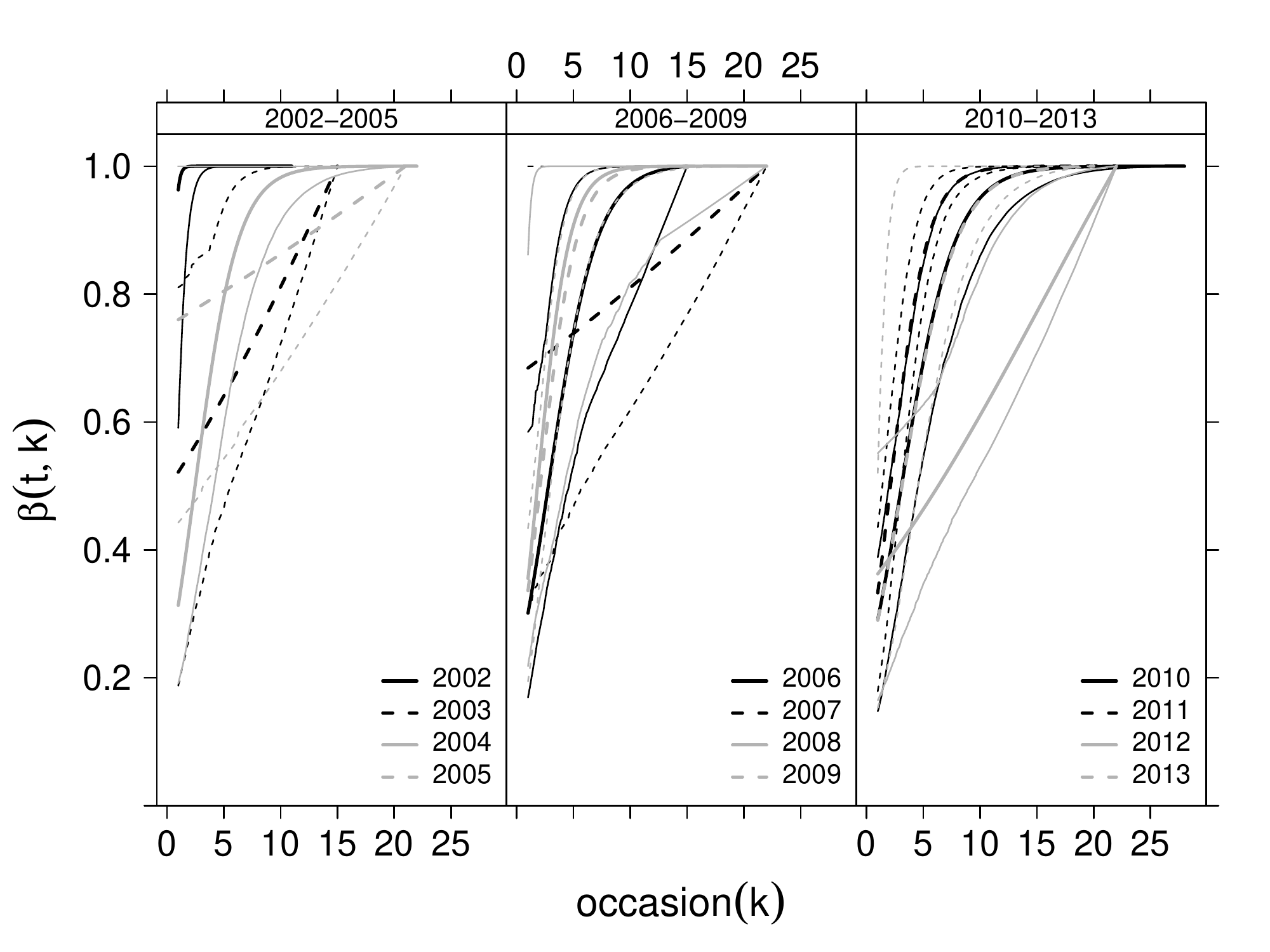}
			\includegraphics[width=0.45\textwidth]{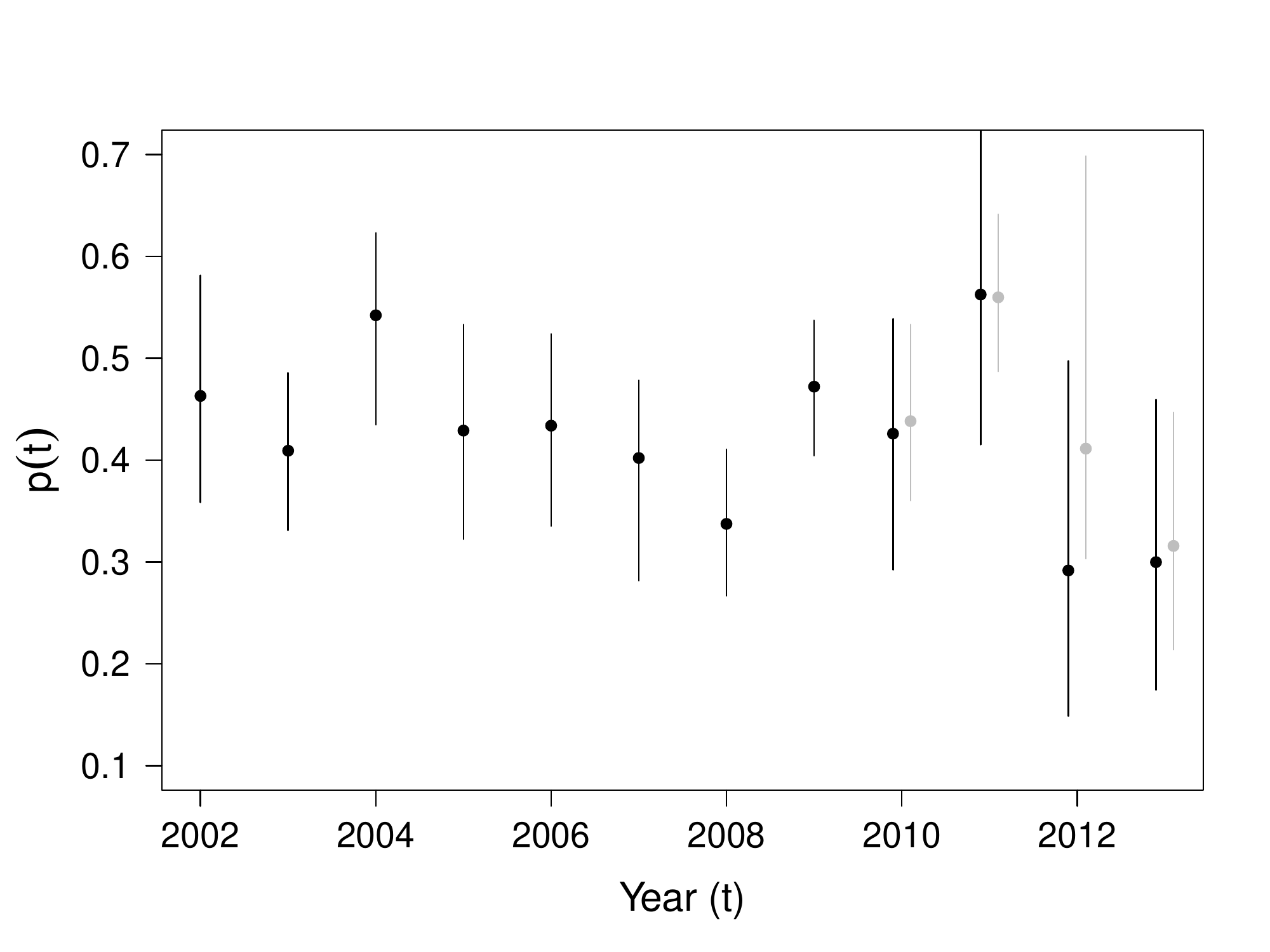}
			\includegraphics[width=0.45\textwidth]{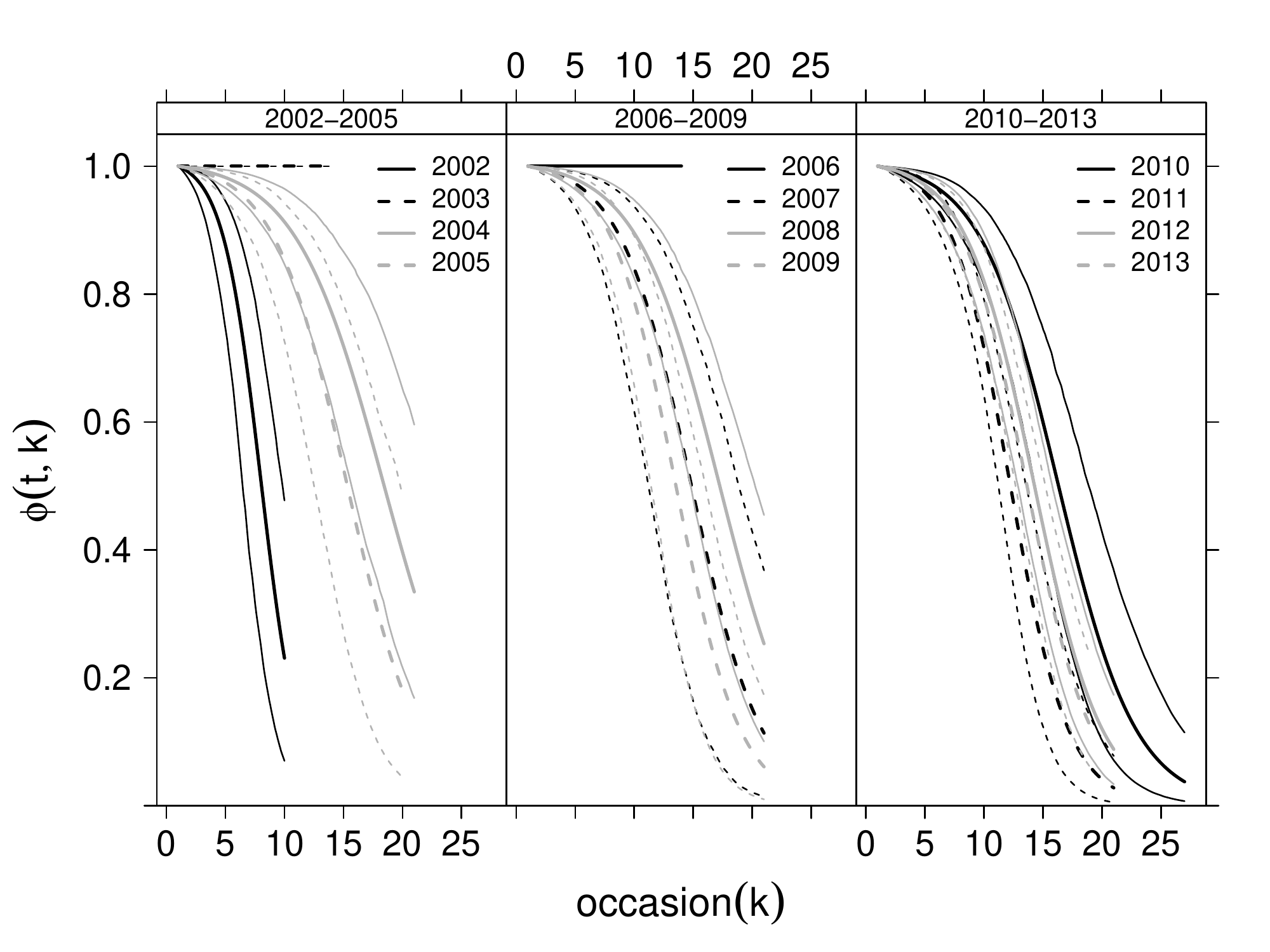}
		\end{center}
		\caption{Maximum likelihood estimates and 95\% bootstrap confidence intervals from the multi-state multi-period stopover model for: (top, left) recruitment probabilities for years 2002-2013; (top, right) arrival probabilities for years 2002-2013; (bottom, left) capture probabilities for years 2002-2013 for the old ponds (black) and years 2010-2013 for the new ponds (grey) and; (bottom, right) retention probabilities for years 2002-2013 of the great crested newt study.}
		\label{fig:newts}
	\end{figure}

	We now consider the additional observable states (old or new ponds).  This additional information is available for the 2010-2013 breeding seasons (all ponds in 2002-2009 are old ponds and so the multi-state parameters are not required for these years).  Due to the large number of capture occasions, and very small population size, we remove the occasion-dependence from the capture probabilities and instead allow them to be dependent on both year and state.  We also estimate the initial discrete state probabilities and transition probabilities between the different observable states for each year where the multi-state data is available (2010-2013).  The results from fitting the multi-state multi-period stopover model are given in Figure~\ref{fig:newts} and Table~\ref{tab:newts}.  Standard errors and 95\% confidence intervals are estimated through a nonparametric bootstrap (resampling individual capture histories).
	
	\begin{table}[!ht]
		\caption{Maximum likelihood estimates and bootstrap standard errors from the multi-state multi-period stopover model for the initial discrete state and transition probabilities for the old (state 1) and new (state 2) ponds for years 2010-2013 of the great crested newt study.}
		\begin{center}
			\begin{tabular}{ccccc}
				& \multicolumn{4}{c}{Year ($t$)}\\
				Parameter & 2010 & 2011 & 2012 & 2013\\ \hline
				$\alpha(t,1)$ & 0.69 (0.11) & 0.28 (0.09) & 0.48 (0.22) & 0.33 (0.10)\\
				$\psi_{12}(t)$ & 0.11 (0.03) & 0.17 (0.05) & 0.26 (0.08) & 0.14 (0.06)\\
				$\psi_{21}(t)$ & 0.06 (0.03) & 0.10 (0.03) & 0.17 (0.10) & 0.10 (0.04)\\ 
			\end{tabular}
		\end{center}
		\label{tab:newts}
	\end{table}
	
	The results indicate the data collection process is close to a complete census of individuals present at the site.  The total population of newts that visited the site at least once during the 12 years of sampling is estimated to be $N=107.2$ (SE 0.43) of which $n=106$ were captured.  The results also indicate some possible differences between the old and the new ponds.  When the new ponds were initially colonised in 2010 there appears to have been a preference for the old ponds as an initial choice when newts arrived at the site.  This is likely due to the amount of vegetation and invertebrates in the ponds; the older ponds had longer to mature than the new ponds and therefore contained significantly more food, predators and plant cover.  It appears that upon arrival at the site the newts have moved towards the new ponds as their initial choice as the new ponds have become more established.  The capture probabilities indicate clear temporal variation and with the exception of 2012 the capture probabilities in both old and new ponds are very similar.  The movement between the ponds is quite low, newts appear to show high fidelity to the type of pond they are in (old or new) with a consistently higher fidelity for the new ponds.  The survival probability for this population of newts between breeding seasons, assumed to be constant between years, is estimated to be 0.82 (SE 0.025).

\section{Discussion} \label{sec:discussion}
	In this paper we have developed a generalised multi-state multi-period stopover model.  This global model for capture-recapture data is an extension to many of the commonly applied capture-recapture models.  In particular the multi-state single-period stopover model combines the AS model and stopover models to allow the capture probabilities to be time-, age- and state-dependent. The new model is a fully open population model able to estimate abundance and therefore likely to resolve long-standing issues concerning the assumption of closure when sampling animal populations repeatedly over short time frames.  The multi-state multi-period stopover model is a further extension of this multi-state stopover model considering multiple periods of capture occasions within a single tractable likelihood.  Models that allow for the combining of information, either across several years of data collection or different sources of information e.g.~count data, are widely used in ecological applications \citep{Besbeas02}.
	
	This likelihood is constructed using an HMM form leading to an efficient likelihood expression that can be maximised using standard optimisation algorithms and software.  This structure also permits the extension to include additional complexities in a straightforward manner.  For example, in this paper we assume that the state information is recorded with certainty. In practice this may not be the case but the model can be extended further to incorporate such state-uncertainty by introducing additional state assignment probabilities \citep{King14b, King16}.  
	
	In these derivations we assume the states are discrete. In the case of continuous state information the approach of the HMMs above could still be applied by using a fine discretization of the continuous states into a discrete form \citep{Langrock13}. Care would need to be adopted in this instance to avoid the dimensions of the matrices involved becoming too large leading to computational issues.
	
	Further extensions to these models could include the addition of a state-dependence to the retention probabilities. This would allow the departure of individuals to be modelled differently depending on their final state in a given year. To reduce the number of parameters estimated from the capture-recapture data alone, covariates could also be considered. As with the multi-period stopover model, consideration could also be given to temporary migration and the idea of individuals skipping attendance in some years. For instance the success or failure to breed in a given year may lead an individual to skip the following year to improve their body condition before returning in later years to reattempt breeding. In the case of the newts this behaviour is more likely in females as they have to invest more energy to produce eggs each year.  This information would need to be incorporated in the primary level of the model where the behaviour in a given year is summarised into a state on the primary level.  Again such extensions can be considered and the efficient HMM likelihood exploited.

\section*{Acknowledgements}
	Financial support was provided by a Carnegie Scholarship awarded to Hannah Worthington by the Carnegie Trust for the Universities of Scotland. Rachel McCrea was funded by a Natural Environment Research Council research fellowship grant NE/J018473/1.  The authors gratefully acknowledge all of the volunteers and students who collected the data from the newts at the University of Kent, particularly Sue Young, Amy Wright and Brett Lewis.

\newpage	
\appendix	

\begin{center}
	\large Supplementary Materials for Estimating abundance from multiple sampling capture-recapture data using hidden Markov models in the presence of discrete-state information
\end{center}
	
\begin{figure}[!htp]
	\begin{center}
		\includegraphics[width=0.45\textwidth]{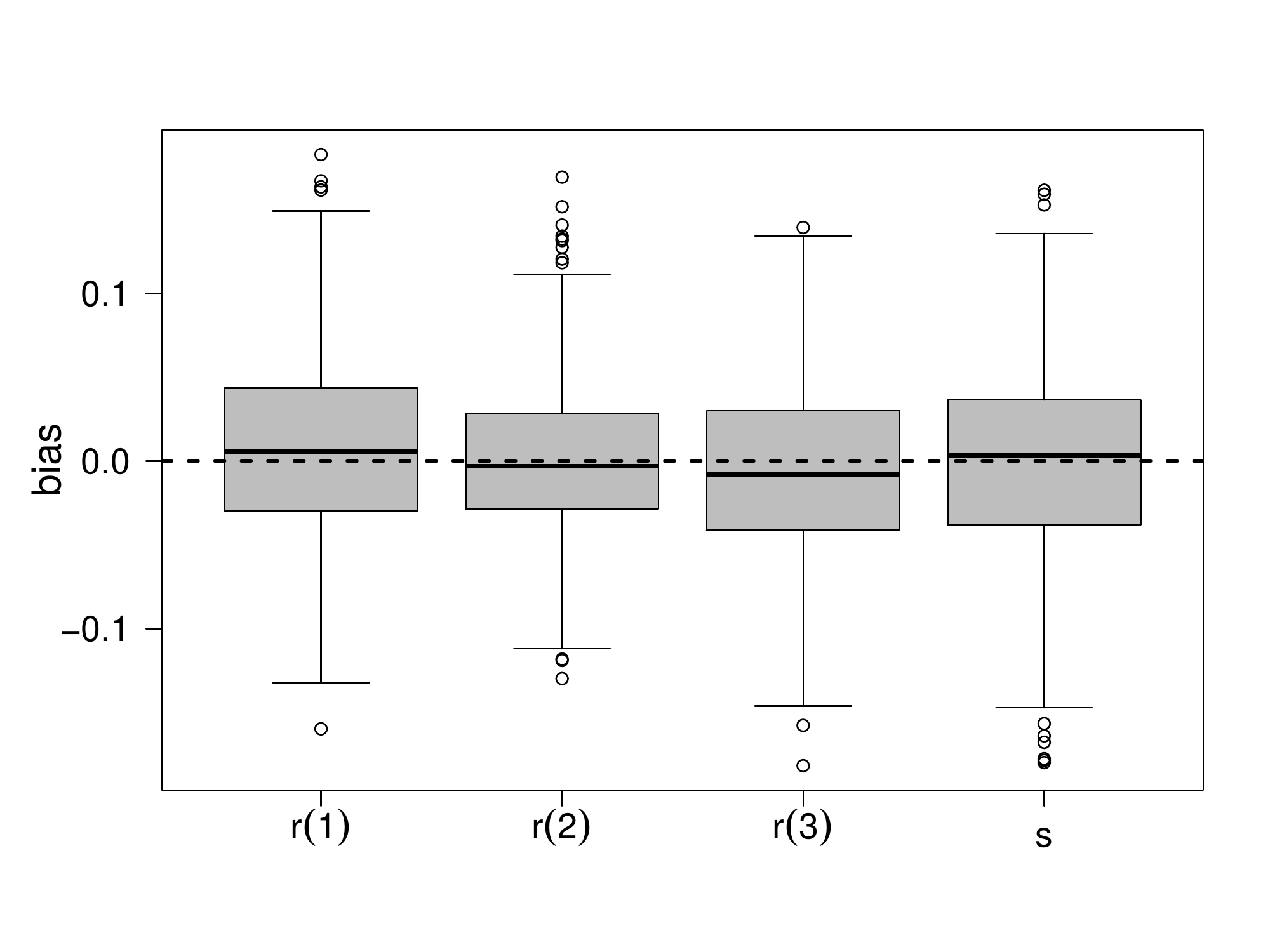}
		\includegraphics[width=0.45\textwidth]{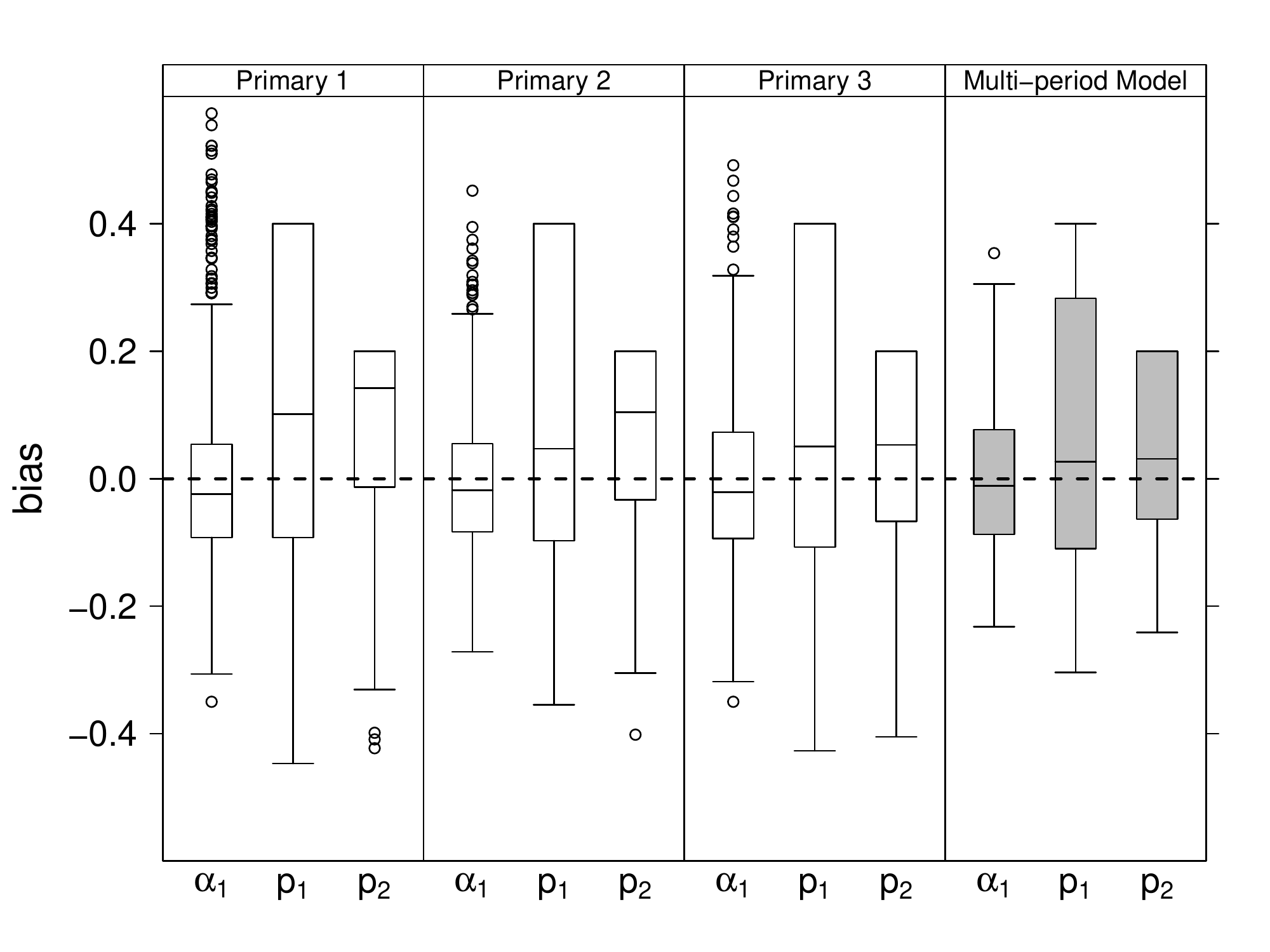}
		\includegraphics[width=0.45\textwidth]{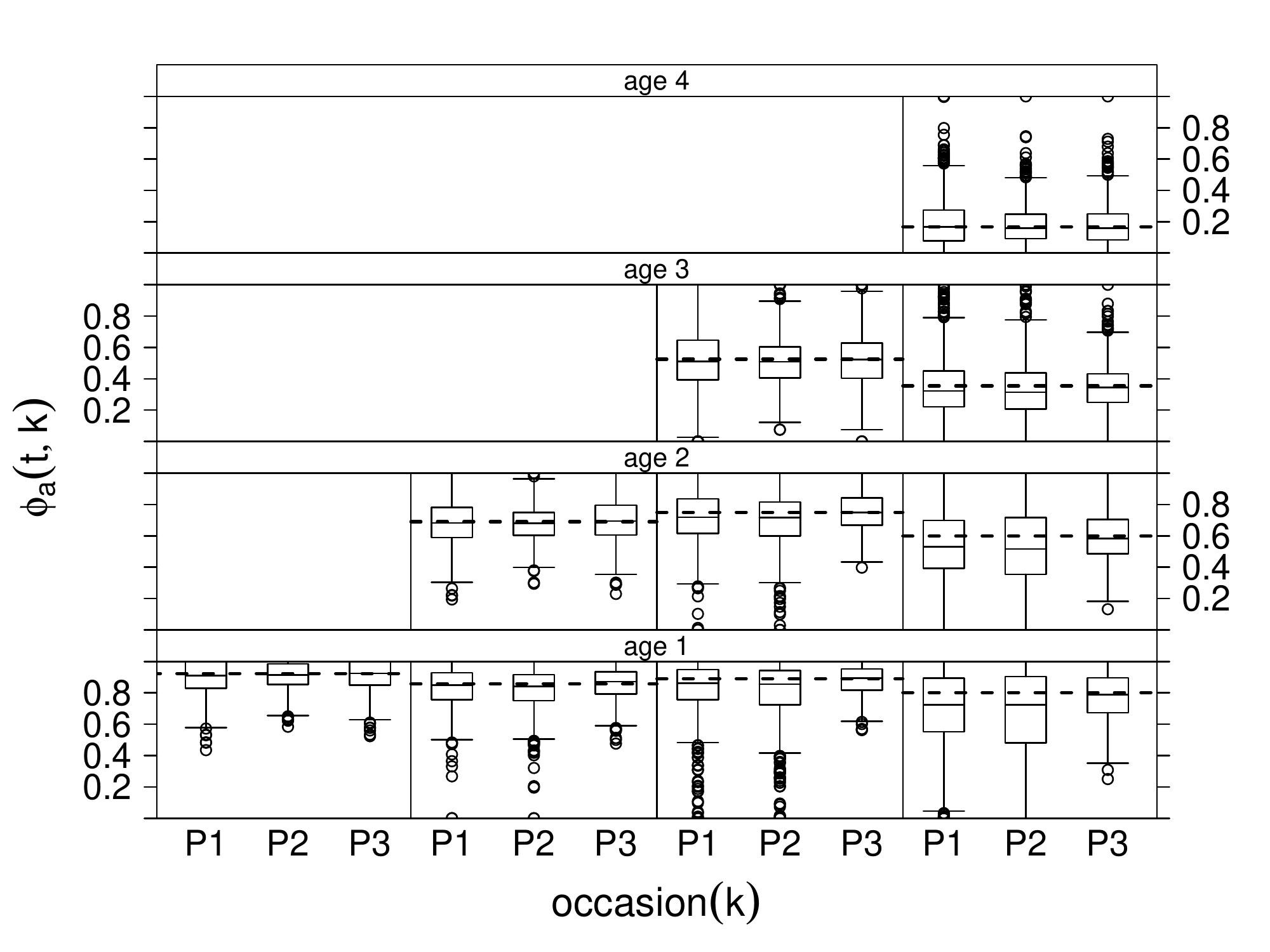}
		\includegraphics[width=0.45\textwidth]{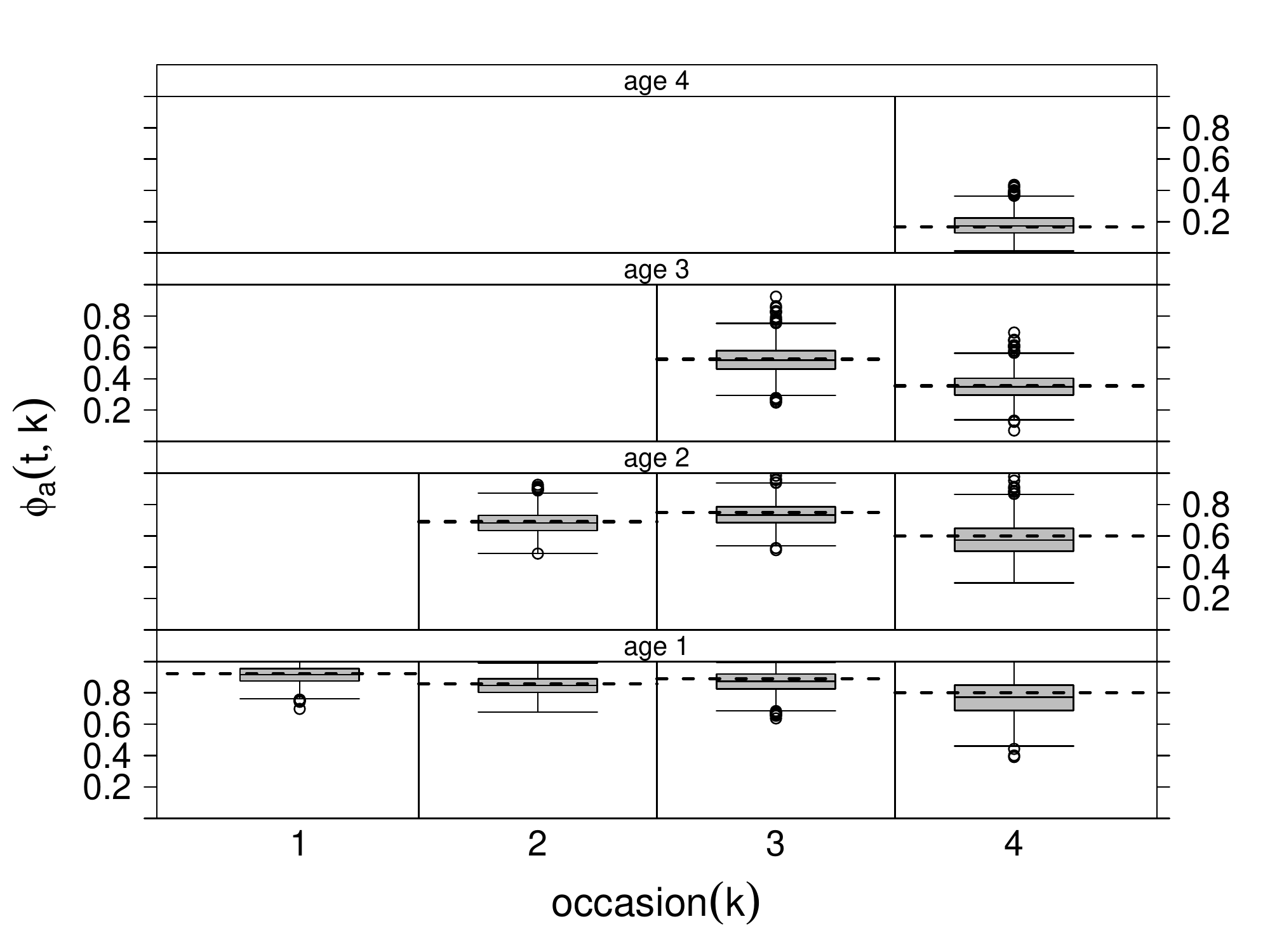}
	\end{center}
	\caption{Results from the simulation study where $N=100$ for: (top, left) bias of the recruitment and survival probabilities for the multi-period model; (top, right) bias of the initial discrete state and capture probabilities in each primary period for the single-period model (white) and the multi-period model (grey); (bottom, left) retention probabilities for each primary period, each capture occasion within primaries and ages for the single-period model; and (bottom, right) retention probabilities for each capture occasion and ages (shared across primaries) for the multi-period model. True parameter values are shown by a dashed line.}
	\end{figure}
	
\begin{figure}[!htp]
	\begin{center}
		\includegraphics[width=0.45\textwidth]{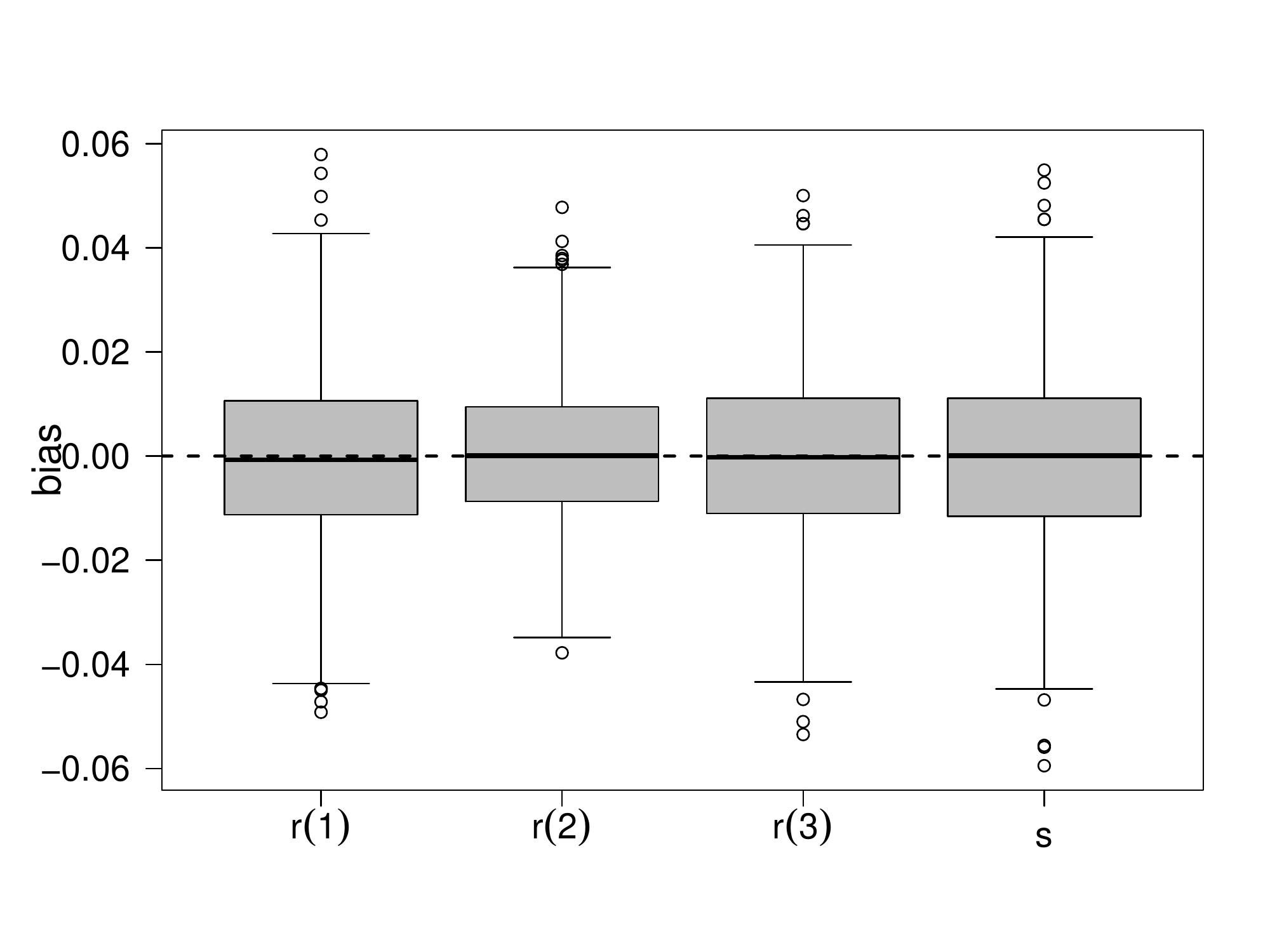}
		\includegraphics[width=0.45\textwidth]{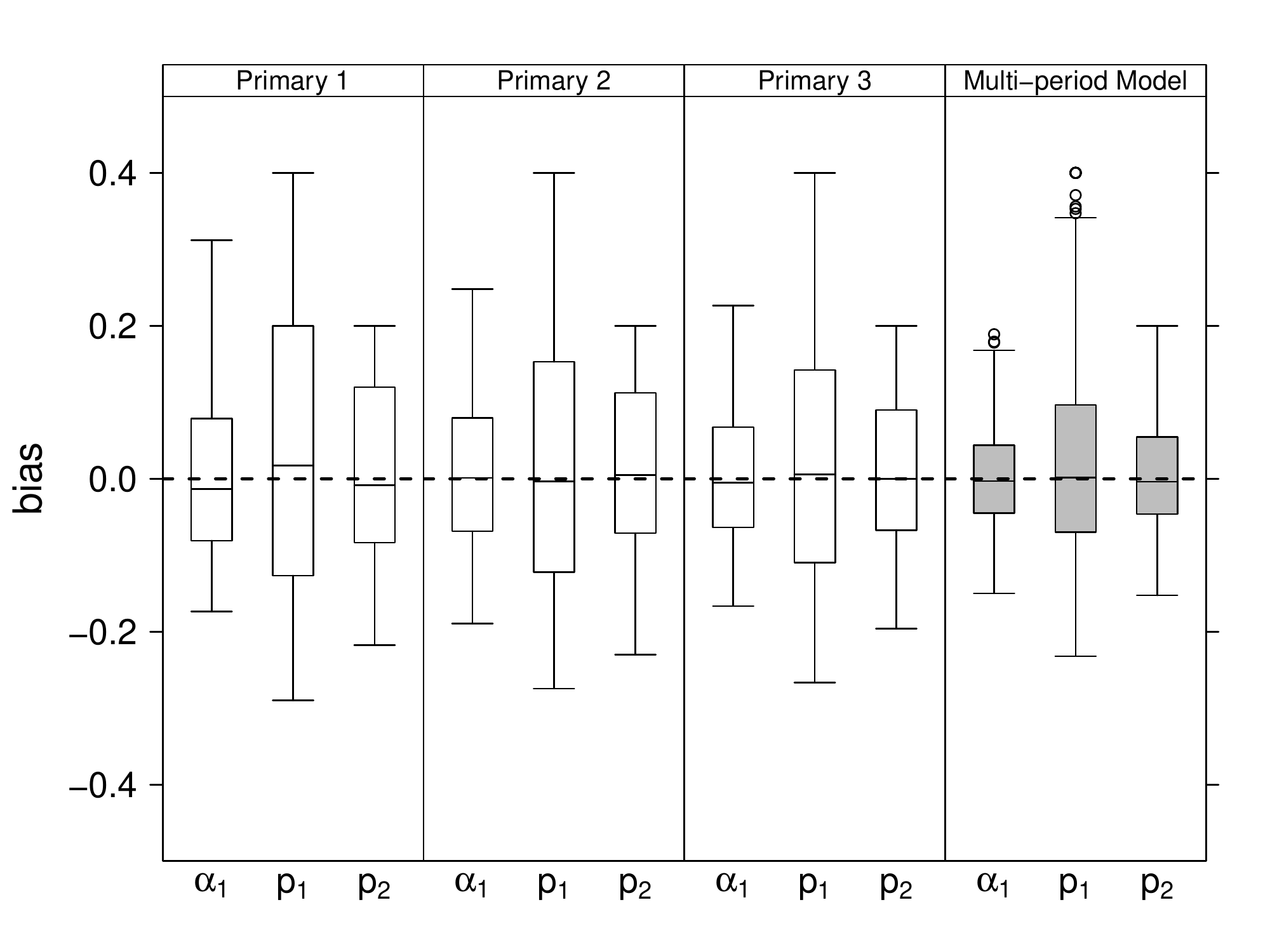}
		\includegraphics[width=0.45\textwidth]{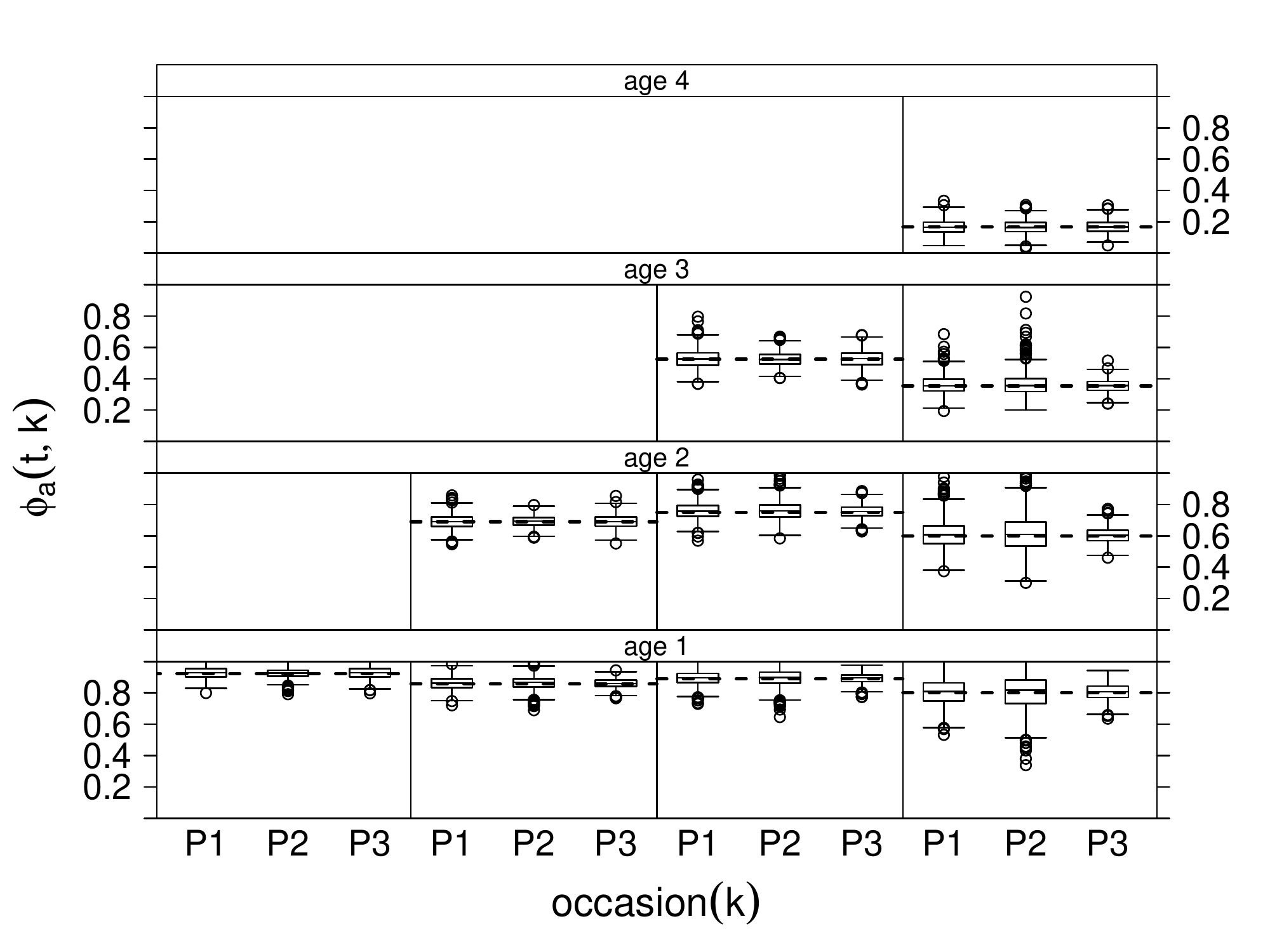}
		\includegraphics[width=0.45\textwidth]{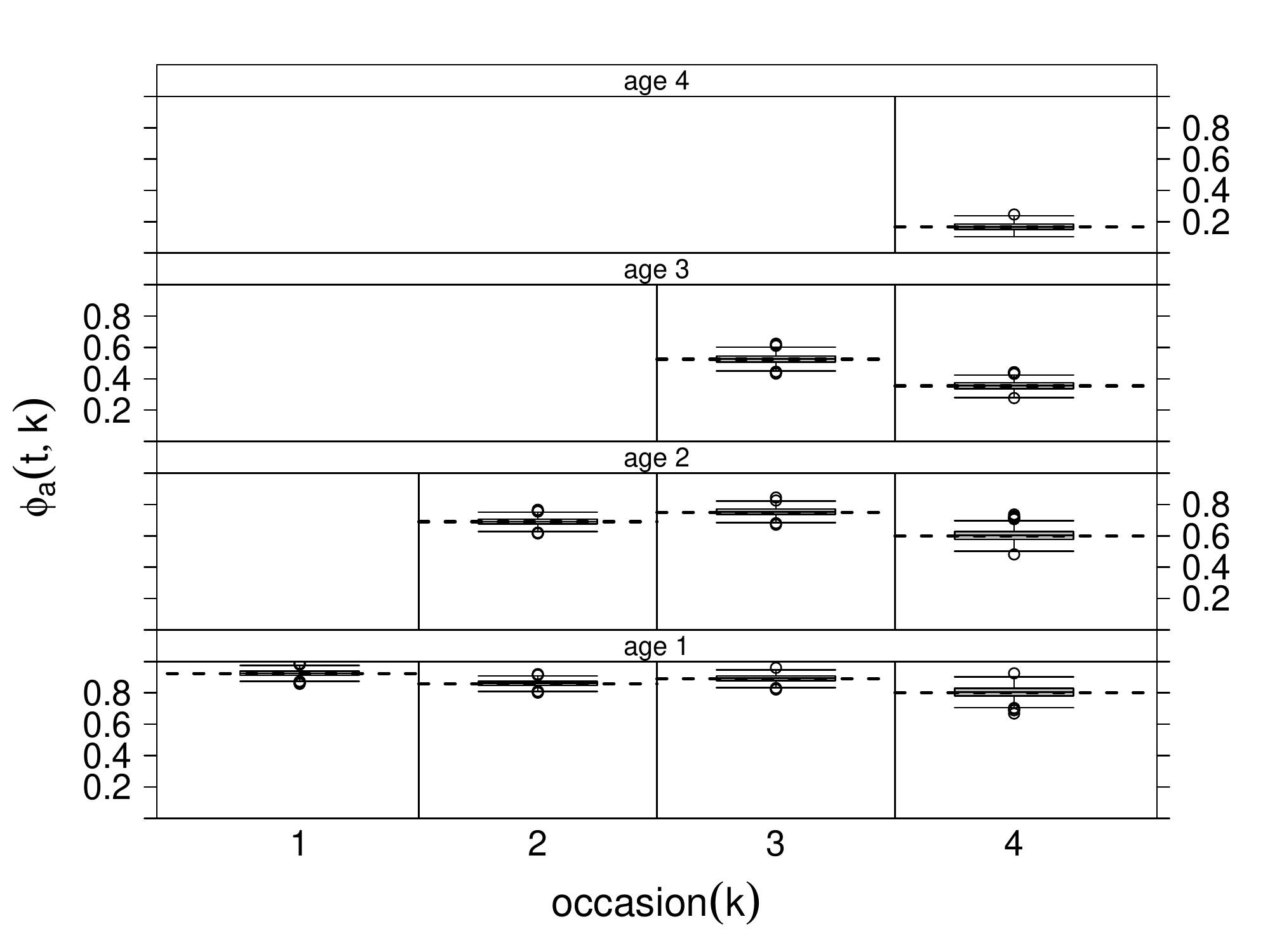}
	\end{center}
	\caption{Results from the simulation study where $N=1000$ for: (top, left) bias of the recruitment and survival probabilities for the multi-period model; (top, right) bias of the initial discrete state and capture probabilities in each primary period for the single-period model (white) and the multi-period model (grey); (bottom, left) retention probabilities for each primary period, each capture occasion within primaries and ages for the single-period model; and (bottom, right) retention probabilities for each capture occasion and ages (shared across primaries) for the multi-period model. True parameter values are shown by a dashed line.}
\end{figure}


\begin{thebibliography}{9}
	\bibitem[\protect\citeauthoryear{Arnason}{Arnason}{1972}]{Arnason72}
	Arnason, A. N. (1972), Parameter estimates from mark-recapture-recovery experiments on two populations subject to migration and death.
	\textit{Researches on Population Ecology} \textbf{13}, 97--113
		
	\bibitem[\protect\citeauthoryear{Arnason}{Arnason}{1973}]{Arnason73}
	Arnason, A. N. (1973), The estimation of population size, migration rates, and survival in a stratified population.
	\textit{Researches on Population Ecology} \textbf{15}, 1--8
		
	\bibitem[\protect\citeauthoryear{Besbeas et al.}{Besbeas et al.}{2002}]{Besbeas02}
	Besbeas, P., Freeman, S. N., Morgan, B. J. T. and Catchpole, E. A. (2002), Integrating mark-recapture-recovery and census data to estimate animal abundance and demographic parameters.
	\textit{Biometrics} \textbf{58}, 540--547
		
	\bibitem[\protect\citeauthoryear{Brownie et al.}{Brownie et al.}{1993}]{Brownie93}
	Brownie, C., Hines, J. E., Nichols, J. D., Pollock, K. H. and Hestbeck, J. B. (1993), Capture-recapture studies for multiple strata including non-Markovian transitions.
	\textit{Biometrics} \textbf{49}, 1173--1187
		
	\bibitem[\protect\citeauthoryear{Cormack}{Cormack}{1964}]{Cormack64}
	Cormack, R. M. (1964), Estimates of survival from the sighting of marked animals.
	\textit{Biometrika} \textbf{51}, 429--438
		
	\bibitem[\protect\citeauthoryear{Dupuis and Schwarz}{Dupuis and Schwarz}{2007}]{Dupuis07}
	Dupuis, J. A. and Schwarz, C. J. (2007), A Bayesian approach to the multistate Jolly-Seber capture-recapture model.
	\textit{Biometrics} \textbf{63}, 1015--1022
		
	\bibitem[\protect\citeauthoryear{Gimenez et al.}{Gimenez et al.}{2007}]{Gimenez07}
	Gimenez, O., Rossi, V., Choquet, R., Dehais, C., Doris, B., Varella, H., Vila, J.-P. and Pradel, R. (2007), State-space modelling of data on marked individuals.
	\textit{Ecological Modelling} \textbf{206}, 431--438
		
	\bibitem[\protect\citeauthoryear{Griffiths et al.}{Griffiths et al.}{2015}]{Griffiths15}
	Griffiths, R. A., Foster, J., Wilkinson, J. W. and Sewell, D. (2015), Science, statistics and surveys: a herpetological perspective.
	\textit{Journal of Applied Ecology} \textbf{52}, 1413--1417
		
	\bibitem[\protect\citeauthoryear{Jolly}{Jolly}{1965}]{Jolly65}
	Jolly, G. M. (1965), Explicit estimates from capture-recapture data with both death and immigration-stochastic model.
	\textit{Biometrika} \textbf{52}, 225--247
		
	\bibitem[\protect\citeauthoryear{Kendall and Bjorkland}{Kendall and Bjorkland}{2001}]{Kendall01}
	Kendall, W. L. and Bjorkland, R. (2001), Using open robust design models to estimate temporary emigration from capture-recapture data.
	\textit{Biometrics} \textbf{57}, 1113--1122
		
	\bibitem[\protect\citeauthoryear{Kendall et al.}{Kendall et al.}{1995}]{Kendall95}
	Kendall, W. L., Pollock, K. H. and Brownie, C. (1995), A likelihood-based approach to capture-recapture estimation of demographic parameters under the robust design.
	\textit{Biometrics} \textbf{51}, 293--308
		
	\bibitem[\protect\citeauthoryear{King}{King}{2012}]{King12}
	King, R. (2012), A review of Bayesian state-space modelling of capture-recapture-recovery data.
	\textit{Interface Focus} \textbf{2}, 190--204
		
	\bibitem[\protect\citeauthoryear{King}{King}{2014}]{King14a}
	King, R. (2014), Statistical Ecology. 
	\textit{Annual Review of Statistics and its Application} \textbf{1}, 401--426
		
	\bibitem[\protect\citeauthoryear{King and Brooks}{King and Brooks}{2003}]{King03}
	King, R. and Brooks, S. P. (2003), Closed form likelihoods for Arnason-Schwarz models.
	\textit{Biometrika} \textbf{90}, 435--444
		
	\bibitem[\protect\citeauthoryear{King and Langrock}{King and Langrock}{2016}]{King16}
	King, R. and Langrock, R. (2016), Semi-Markov Arnason-Schwarz models.
	\textit{Biometrics} \textbf{72}, 619--628
		
	\bibitem[\protect\citeauthoryear{King and McCrea}{King and McCrea}{2014}]{King14b}
	King, R. and McCrea, R. S. (2014), A generalised likelihood framework for partially observed capture-recapture-recovery models.
	\textit{Statistical Methodology} \textbf{17}, 30--45
		
	\bibitem[\protect\citeauthoryear{King et al.}{King et al.}{2009}]{King09}
	King, R., Morgan, B. J. T., Gimenez, O. and Brooks, S. P. (2009), \textit{Bayesian analysis for population ecology}.
	CRC Press, Boca Raton.
		
	\bibitem[\protect\citeauthoryear{Langrock and King}{Langrock and King}{2013}]{Langrock13}
	Langrock, R. and King, R. (2013), Maximum likelihood estimation of mark-recapture-recovery models in the presence of continuous covariates.
	\textit{Annals of Applied Statistics} \textbf{7}, 1709--1732
		
	\bibitem[\protect\citeauthoryear{Lebreton et al.}{Lebreton et al.}{2009}]{Lebreton09}
	Lebreton, J. D., Almeras, T. and Pradel, R. (2009), Competing events, mixtures of information and multistratum recapture models.
	\textit{Bird Study} \textbf{46}, S39--S46
		
	\bibitem[\protect\citeauthoryear{Lewis et al.}{Lewis et al.}{2017}]{Lewis17}
	Lewis, B., Griffiths, R. A. and Wilkinson, J. W. (2017), Population status of great crested newts (\textit{Triturus cristatus}) at sites subjected to development mitigation.
	\textit{Herpetological Journal} \textbf{27}, 133--142
		
	\bibitem[\protect\citeauthoryear{McCrea and Morgan}{McCrea and Morgan}{2011}]{McCrea11}
	McCrea, R. S. and Morgan, B. J. T. (2011), Multistate mark-recapture model selection using score tests.
	\textit{Biometrics} \textbf{67}, 234--241
		
	\bibitem[\protect\citeauthoryear{McCrea et al.}{McCrea et al.}{2010}]{McCrea10}
	McCrea, R. S., Morgan, B. J. T., Gimenez, O., Besbeas, P., Lebreton, J. -D. and Bregnballe, T. (2010), Multi-Site Integrated Population Modelling.
	\textit{Journal of Agricultural, Biological and Environmental Statistics} \textbf{15}, 539--561
		
	\bibitem[\protect\citeauthoryear{Pledger et al.}{Pledger et al.}{2009}]{Pledger09}
	Pledger, S., Efford, M., Pollock, K. H., Collazo, J. A. and Lyons, J. E. (2009), Stopover duration analysis with departure probability dependent on unknown time since arrival.
	\textit{Environmental and Ecological Statistics (Edited by D. L. Thomson, E. G. Cooch and M. J. Conroy)} \textbf{3}, 349--363
		
	\bibitem[\protect\citeauthoryear{Pollock}{Pollock}{1982}]{Pollock82}
	Pollock, K. H. (1982), A capture-recapture design robust to unequal probability of capture.
	\textit{The Journal of Wildlife Management} \textbf{46}, 752--757
		
	\bibitem[\protect\citeauthoryear{Royle}{Royle}{2008}]{Royle08}
	Royle, J. A. (2008), Modeling individual effects in the Cormack-Jolly-Seber model: a state-space formulation.
	\textit{Biometrics} \textbf{64}, 364--370
		
	\bibitem[\protect\citeauthoryear{Schofield and Barker}{Schofield and Barker}{2008}]{Schofield08}
	Schofield, M. R. and Barker, R. J. (2008), A unified capture-recapture framework.
	\textit{Journal of Agricultural, Biological and Environmental Statistics} \textbf{13}, 459--477
		
	\bibitem[\protect\citeauthoryear{Schwarz and Arnason}{Schwarz and Arnason}{1996}]{Schwarz96}
	Schwarz, C. J. and Arnason, A. N. (1996), A general methodology for the analysis of capture-recapture experiments in open populations.
	\textit{Biometrics} \textbf{52}, 860--873
		
	\bibitem[\protect\citeauthoryear{Schwarz et al.}{Schwarz et al.}{1993}]{Schwarz93}
	Schwarz, C. J., Schweigert, J. F. and Arnason, A. N. (1993), Estimating migration rates using tag recovery data.
	\textit{Biometrics} \textbf{59}, 291--318
		
	\bibitem[\protect\citeauthoryear{Seber}{Seber}{1965}]{Seber65}
	Seber, G. A. F. (1965), A note on the multiple-recapture census.
	\textit{Biometrika} \textbf{52}, 249--259
		
	\bibitem[\protect\citeauthoryear{Zucchini et al.}{Zucchini et al.}{2016}]{Zucchini16}
	Zucchini, W., MacDonald, I. L. and Langrock, R. (2016), \textit{Hidden Markov Models for Time Series - An Introduction Using R (2nd Edition)}.
	Chapman and Hall/CRC.
\end{thebibliography}
\end{document}